# Geosynchronous magnetopause crossings: necessary conditions

A. Suvorova[1,2], A. Dmitriev[1,2], J.-K. Chao[1], M. Thomsen[3], Y.-H. Yang[1]

[1]Institute of Space Science, National Central University, Taiwan
[2]Skobeltsyn Institute of Nuclear Physics, Moscow State University, Russia
[3]Los Alamos National Laboratory, USA

A. V. Suvorova, Institute of Space Science National Central University, Chung-Li, 320, Taiwan; also Skobeltsyn Institute of Nuclear Physics, Moscow State University, Russia (e-mail: suvorova_alla@yahoo.com)

A. V. Dmitriev, Institute of Space Science National Central University, Chung-Li, 320, Taiwan (e-mail: dalex@jupiter.ss.ncu.edu.tw)

J. K. Chao, Institute of Space Science National Central University, Chung-Li, 320, Taiwan (e-mail: jkchao@jupiter.ss.ncu.edu.tw)

M. F. Thomsen, Space and Atmospheric Sciences (NIS-1) MS D466 Los Alamos National Laboratory, Los Alamos, NM 87545 (e-mail: mthomsen@lanl.gov)

Y.H. Yang, Institute of Space Science National Central University, Chung-Li, 320, Taiwan (e-mail: yhyang@jupiter.ss.ncu.edu.tw)

**Abstract.** The experimental data on GOES magnetic measurements and plasma measurements on LANL geosynchronous satellites is used for selection of 169 case events containing 638 geosynchronous magnetopause crossings (GMCs) in 1995 to 2001. The GMCs and magnetosheath intervals associated with them are identified using advanced methodic that take into account: 1. Strong deviation of the magnetic field measured by GOES from the magnetospheric field; 2. High correlation between the GOES magnetic field and interplanetary magnetic field (IMF); 3. Substantial increase of the low energy ion and electron contents measured by LANL. Accurate determination of the upstream solar wind conditions for the GMCs is performed using direct solar wind propagation from an upstream monitor (Wind, Geotail, ACE) to the geosynchronous orbit as well as additional independent criteria such as variations of geomagnetic activity (*Dst* (SYM) index), correlation of the magnetic field measured by GOES with IMF, and correlation of low energy plasma density in the magnetopsheath with upstream solar wind plasma density. The location of the GMCs and associated upstream solar wind conditions are ordered in aberrated GSM coordinate system with X-axis directed along the solar wind flow. In the selected data set of GMCs the solar wind total pressure *Psw* varies up to 100 nPa and southward IMF *Bz* achieves 60 nT. We study the necessary conditions for the geosynchronous magnetopause crossings using scatter plot of the GMCs in the coordinate space of *Psw* versus *Bz*. In such representation the upstream solar wind conditions demonstrate sharp envelope boundary under which no GMCs are occurred. The boundary has two strait horizontal branches where *Bz* does not influence on the magnetopause location. The first branch is located in the range of *Psw*=21 nPa for large positive *Bz* and is associated with an asymptotic regime of the pressure balance. The second branch asymptotically approaches to the range of *Psw*=4.8 nPa under very strong negative *Bz* and it is associated with a regime of the *Bz* influence saturation. We suggest that the saturation is caused by relatively high contribution of the magnetosphere thermal pressure into the pressure balance on the magnetopause. The intermediate region of the boundary for the moderate negative and small positive IMF *Bz* can be well approximated by a hyperbolic tangent function. We interpret the envelope boundary as a range of necessary upstream solar wind conditions required for GMC in the point on the magnetopause located mostly close to the Earth ("perigee" point). We obtain that the dipole tilt angle and dawn-dusk asymmetry influence on the "perigee" point location. We find that the aGSM latitude of this point depends linearly on the dipole tilt angle with the slope about –0.5. The aGSM longitude of the "perigee" point decreases with IMF *Bz* with a rate of about 2 angular minutes per 1 nT. An empirical model predicting the magnetopause crossing of the geosynchronous orbit in the "perigee" point is proposed.

# 1. Introduction

Present paper is a collection of results published in a number of papers [*Dmitriev and Suvorova* 2004; *Dmitriev et al.*, 2004, 2005a,b, 2010; 2011; *Suvorova et al.*, 2005; *Karimabadi et al.*, 2007]. In those papers, such important effects as magnetosphere dawn-dusk asymmetry, reconnection saturation, large-scale magnetopause undulations and others were studied on the base of magnetopause crossings observed by geosynchronous satellites. The original paper has been submitted to a peer-reviewing journal in 2003 [*Suvorova et al.*, 2005]. However, after revision a part, containing physical interpretations of our findings, has been totally removed because of strong resistance of a Reviewer. Later, those interpretations and suggestions have been proved and published in different papers.

During last years, an interest to effects of extreme solar wind (SW) conditions to the magnetosphere size and shape continuously grows up. Among a group of ten of magnetopause (MP) empirical models created last decade a half are able to predict MP location at dayside under extreme solar wind (SW) conditions. A complete list of them with improved versions is constituted from [*Shue et al.,* 2000b, *Suvorova et al.,* 1999; *Yang et al.*, 2002] and contains following models: *Petrinec and Russell* [1996] (PR96), *Shue et al.* [1997,1998] (Sh98), *Kuznetsov and Suvorova* [1998a] and *Kuznetsov et al.* [1998] (KS98), *Dmitriev et al.* [1999] and *Dmitriev and Suvorova* [2000] (DS00), *Chao et al.* [2002] and *Yang et.al*. [2003] (Ch02). Figure 1 presents the solar wind dynamic pressure and IMF *Bz* predicted for magnetopause subsolar distance 6.6 Earth's radii (Re) by different MP models. As one can see all of the five MP models predict different SW conditions for geosynchronous magnetopause crossings (GMCs). The differences increase with increasing disturbance level of the solar wind parameters. Table 1 indicates values of the SW dynamic pressure predicted by different models for GMCs in the subsolar point for strong northward (*Bz*=30 nT) and southward (*Bz*=-30 nT) IMF.

For the positive *Bz* the MP models demonstrate very wide scatter in prediction of the SW dynamic pressure (up to 2 times) required for the GMCs. The KS98 and Ch02 models give, respectively, the minimal (*Pd*=24.5 nPa) and maximal (*Pd*=45 nPa) values. For the strong southward IMF (*Bz*=-30 nT) the spreading in the model predictions become much wider and varies from $Pd$=0.7 nPa for PR96 model to $Pd$=7.4 nPa for Sh98 model (i.e. more than 10 times). Note that the PR96 model does not include the effect of "Bz influence saturation under strong southward IMF" which is predicted by other four MP models. However even the MP models containing "*Bz* saturation" effect produce different predictions varying from $Pd$~3 nPa to $Pd$~12 nPa. There are two questions arising from Figure 1 and Table 1. What is a reason of notable disagreement between the MP model predictions in the geosynchronous orbit? Which is the mostly realistic MP model?

Various reasons of disagreement between the MP models were comprehensively reviewed in recent papers [*Shue et al.,* 2000 a; 2000 b]. In general, some of them are connected with 1) a quality, 2) a time resolution, 3) a representativeness of data both in geometrical and parameter spaces, 4) a valid choice of an appropriate SW monitor, 5) correct determination of a time delay owing to propagation of the SW structure toward the Earth. The other reasons depend on 6) a choice of mathematical models fitting the MP shape, 7) physical laws that describe basic processes of the MP formation and drive the MP dynamics over whole surface, 8) methodic of data treatment. In the Introduction we only put our attention on databases of the models and their treatment (items 1)-3) and 8)). For items 2)-5) we will make comprehensive analysis in the present study.

Remember that the PR96, Sh98 and Ch02 models were constructed on 5-minute SW data from IMP-8 and ISEE-3 satellites, practically the same basic database of high apogee satellite measurements. The PR96 model did not take into account the time delay on SW propagation from an upstream monitor (IMP-8) to the Earth. In the Sh98 model the time delay was accepted of about 10 and 55 minutes for IMP-8 and ISEE-3 respectively. In the Ch02 model the time delay is calculated using an assumption on direct SW propagation. The rest two models (KS98, DS00) were constructed on hourly averaged SW data and the same basic massive of MP crossings by *Roelof and Sibeck* [1993] with adding a subset from geosynchronous crossings database of [*Kuznetsov and Suvorova,* 1997]. The database is accessible on-line on http://dec1.npi.msu.su/~alla/mp/gmc.html.

Now we turn to the paper sources and give some comments about the results of previous comparison analysis of the MP models. Note, that the improved modified version of the CH02 [*Yang et al.,* 2003] is the latest MP model and it is never yet being compared anywhere. The model DS00 also was not compared with more recent models, though it is an unique multi-parametric model based on perspective method of artificial neural network which permits selecting the best set of input magnetopause model parameters [*Dmitriev and Orlov,* 1997; *Dmitriev et al.,* 1999; *Dmitriev and Suvorova* 2000a; 2000b]. The latest two studies by *Shue et al.* [2000b; 2001] were devoted to detail comparison of the PR96, SH98, KS98 models under extreme SW condition with in situ observations. The ability to predict the MP location near geosynchronous orbit was evaluated. Conclusions of two papers are different concerning to the KS98 and PR96 models, especially many inconsistent explanations were given for the KS98. Fortunately, it is easier for us to clear this strange situation owing to one of us was the coauthor of the KS98 model. We give here more clear and important comments on the used data set of crossings and its treatment in KS98 model, though some references on the analysis of data was included in reference list of [*Kuznetsov and Suvorova,* 1998a]

So, *Shue et al.* [2000b] have obtained that prediction capabilities of two models, SH98 and PR96, are the best, and because of expected the highest false alarm rate (FAR), KS98 model was not compared with observations. Consequently, according to *Shue et al.* [2000b] the KS98 model is worse then two other models, while according to *Shue et al*. [2001] the KS98 is one of the best due to high probability of correct prediction (PCP). A main root of the contradictions lies in a quality of database. Data set used in KS98 model was mostly criticized [*Shue et al.,* 1998; 2000a; 2000b]. The most serious remarks on database quality are following: not a homogeneous of *Roelof and Sibeck's* data set and hourly averaged SW parameters. It is very hard to argue with these arguments. Instead this we discuss meaning and effect of these remarks in connection to data treatment used for a model developing.

Firstly, note that the KS98 is indeed developed on highly inhomogeneous data, which also includes 39 GMCs. Undoubtedly inhomogeneous data have some missadvantages indicated in [*Shue et.al,* 1997,2000a]. However, the data accumulated using different experiments and operated by different authors is practically free from any systematical errors. Moreover, an investigator working with low quality data usually cleans the massive. When a data set is binned into small intervals of model parameters (space and physical) the spread of point is checked. If the MP distance for one of the points is unreal, taking into account averaged SW conditions and spreading of the neighbor points, then this point is excluded from further consideration. Such procedure was performed in the KS98 model.

Secondly, non-homogeneity of the experimental data has much deeper meaning. The dynamics of the dayside MP, especially of the subsolar point, is different from the flank and tail MP regions [*Kuznetsov et al.*, 1992; *Kuznetsov and Suvorova*, 1994; *Petrinec and Russell*, 1996]. Hence, even data on magnetopause crossings selected in different regions of the magnetopause by a single author using similar onboard detectors is inhomogeneous in sense of driving parameters. In other words, fitting over all MP surface (including dayside, flank and tail regions) should add large errors in determination of the dynamics of the subsolar point where number of the MP crossings is usually much smaller than in the other regions. Due to this a dependence of the magnetosphere size at the subsolar point (under a simpler assumption about the axially symmetric MP shape) should be fitted inside a narrow range of angles (say less then 35 degrees) calculated from axis of symmetry. This is one of the reasons why the first version of the MP model by *Kuznetsov and Suvorova* [1993; 1996a] is significantly different from the model by *Roelof and Sibeck* [1993] despite these models are developed using absolutely the same data set. Moreover, *Kuznetsov and Suvorova* [1993, 1996a] reveal in the range of $B_z$ from $-10$ nT to $-6$ nT weakening of the negative $B_z$ influence and a clear tendency of turning at "saturation" regime (figure 4 in the original paper). This is a good example of importance of the data treatment in the MP model development. An approach in data treatment used in KS98 model is very similar to ANN operation, and apropos ANN is the most effective method to analyze just low quality data with many gaps and spikes.

Thirdly, a strong argument about using hourly averaged SW parameters is discussed in connection with the remark on using GMC's in modeling. In review papers [*Shue et al.*, 2000a; 2000b] the authors caution model that including the geosynchronous crossings could add orbital bias to the data set and the orbital bias is the most important issue affecting the quality of a data set. In general the statement is true.

The orbital bias should lead to overestimation of the MP distance, so real MP should be inside the geosynchronous orbit for the most cases of large pressure and/or large southward *Bz*. This means that modeling based on all occurred (i.e. any) GMCs should inevitably overestimate the SW conditions required for the MP location at 6.6 Re. Consideration of Figure 1 and Table 1 shows clearly that practically always the KS98 model "suffering on the orbital bias" predicts weaker solar wind conditions (smaller dynamic pressure and less negative *Bz*) than other MP models. That is to say instead overestimation of the SW conditions the KS98 model rather underestimate them, if we accept that the models without orbital bias (PR96, Sh98) do not overestimate the SW conditions. On the other hand comparison of the MP models in [*Shue et al.*, 2001] reveales that the probability of correct prediction (PCP) for the KS98 model as high as for PR96 and SH98 models. It means that the KS98 model can not significantly underestimate the SW conditions because in opposite case the PCP falls down. Does these disagreements indicate that actual SW conditions required for GMC are significantly weaker than predicted by the weakest of the models (i.e. KS98, or PR96, or DS00 (By=20)), that is all models overestimate? No, because in this case the interplanetary medium at geosynchronous orbit should be recognized in many and many cases, but in reality it is not so [*Shue et al.*, 2001; *Yang et al.*, 2002]. Summarizing above discussion we can say that different studies provide us inconsistent conclusions about validity of MP models. Nevertheless from above discussion we can certainly say that the KS98 model can not overestimate the SW conditions for GMC, i.e. the model is free from the orbital bias effect.

The KS98 model has avoided the orbital bias effect due to two circumstances: hourly averaged upstream solar wind data and application of special methodic for minimization of the SW conditions required for GMC. *Shue et al.* [1998] criticize the results of the MP models developed on base of *Roelof and Sibeck* [1993] data set because using the hourly averaged solar wind data tends to underestimation the peak values of the SW parameters. However, such underestimation partially compensates the orbit bias effect causing overestimation of the SW conditions. Moreover, quasi-stationary (within 30 minute - 1 hour) SW conditions is widely used requirement in selection of the MP crossings. When the SW conditions change strongly and fast the equilibrium location of the MP can be very far from the satellite observing the crossing and such events are usually excluded from analysis. In the interplanetary medium the quasi-stationary conditions causing the GMCs are mostly observed inside the ICME [*Dmitriev et al.,* 2005c] where the character time of the IMF rotation is about several hours, dynamic pressure is moderate and varies rather slightly than dramatically. When the IMF *Bz* and solar wind pressure change gradually inside the ICME the time resolution does not play crucial role for determination of the upstream SW conditions. In our study we find many cases characterized by the quasi-stationary solar wind conditions mentioned above, i.e. with moderate SW pressure (less than 7 nPa) and large negative *Bz* (less than -10 nT).

A special methodic developed by *Kuznetsov and Suvorova* [1997; 1998b] for selection of the GMCs is based on determination of a surface of minimal conditions required for GMC in three dimensional space of the SW dynamic pressure, IMF *Bz* and local time. For every value of the IMF *Bz* and local time the lowest value of the SW dynamic pressure is searched. It was reasonably assumed that the SW conditions selected in such way are necessary for equilibrium MP to locate just near the geosynchronous orbit. To satisfy the requirement of quasi-stationary SW conditions the short-duration (<6 min) GMCs were excluded from consideration. Thus, the method permits to reveal intervals with quasi-stationary SW conditions under which the MP is located mostly close to the geosynchronous satellite. A large initial statistics of 197 GMCs and wide dynamic range of the SW parameters (pressure varies from 2 nPa to 50 nPa, *Bz* varies from –28 nT to +20 nT) ensured a reasonable result. After a selection only 39 CMC's (including 12 crossing under northward *Bz*) from initial data set were used to develop the KS98 model. More details about the method readers can find in [*Suvorova et al.,* 1999; *Dmitriev and Suvorova,* 2000]. Hence advantages in data treatment provide that the KS98 model has high PCP as well as PR96 and Sh98 despite the quality of the database and hourly averaged SW parameters in comparing with homogenous database and high time resolution of the upstream SW data in two other models.

Relatively weak solar wind conditions required by the PR96, KS98 and DS00 models for the GMCs may lead to frequent false predictions of magnetopause crossings on geosynchronous orbit. The false alarm rate, FAR, is compared by *Shue et al.* [2000b] for PR96, KS98 and Sh98 models. The FAR is revealed high for all compared MP models despite significant difference in their requirements on the SW pressure for GMC. To explain this result *Shue et al.* [2000b] have pointed out that the preconditioning of the IMF *Bz* should be also important. From our point of view the high FAR is owing to different reasons for different models. These reasons connect with many factors that strongly affect the capability of prediction, such as the actual location of the geosynchronous satellite in the GSM system [*Degtyarev et al.*, 1985; *Grafodatskii et al.*, 1989], dipole tilt angle effect [*Petrinec and Russell,* 1995; *Boardsen et al.*, 2000], effect of IMF By [*Dmitriev and Suvorova*, 2000], dawn-dusk asymmetry [*Rufenach et al.*, 1989; *Itoh and Araki*, 1996; *Kuznetsov and Suvorova,* 1996b,1997,1998b] and so on. Moreover the FAR as well as PCP are critically dependent on correct choice of a time delay for the solar wind propagation from upstream monitor to the Earth's magnetosphere because any shift of the upstream parameter time profile leads to decrease PCP and increase FAR. This problem is still sharp and is discussed in the literature intensively [e.g. *Collier et al.*, 1998; *Richardson and Paularena*, 2001; *Weimer et al.,* 2002]. Hence it seems that preconditioning by the IMF *Bz* suggested by *Shue et al.* [2000b] should be considered only after accounting all these well-known effects.

In studies devoted to comparison of the MP models we do not find any comments on the model predictions under northward IMF. However, nobody cancel the pressure balance on magnetopause for positive *Bz*. It is easy to estimate from the pressure balance equation the theoretical value for the SW pressure in the MP subsolar point at distance 6.6 Re. According to [*Spreiter et al.*,1966; *Schield*, 1969] for the MP subsolar point:

$$kP_{SW} = \frac{(2fB)^2}{8\pi} \qquad (1.1)$$

where the coefficient *f* is equal to a ratio of the geomagnetic field strength at the MP to double dipolar value and accounts for a contribution of the large-scale magnetospheric current system to the magnetic field on the MP. The coefficient *k* is equal to the ratio of the magnetosheath plasma pressure applied to the MP to the SW dynamic pressure in the interplanetary space [*Spreiter et al.*, 1966]. The coefficient *k* must always be <1 [*Spreiter et al.*, 1966], while the coefficient *f* depends on the position of the point on the MP and can be >1 [*Schield*, 1969].

For the nominal solar wind conditions, empirical values of $f^2/k$ near the subsolar region are respectively 1.66±0.53 and 1.44±0.29 according to data from OGO-5 [*Holzer and Slavin*, 1978] and PROGNOZ-10 [*Kuznetsov and Suvorova*, 1994]. This fits well with the theoretical value of 1.69 for *f*=1.22, *k*=0.881 obtained by *Spreiter et al.* [1966]. According to [*Kuznetsov and Suvorova*, 1994] an average value of the *f* in the noon sector is 1.02 and it increases up to 2 on the flank of MP. It is rather difficult to determine empirically the value for the coefficient *k*. Thus, only theoretical values for *k* from various models of the SW are known. One of them obtained by *Spreiter et al.*, [1966] from gasdynamic approach is equal to 0.881. Hence we can estimate from (1.1) the dynamic pressure *Psw* that is required for the MP subsolar point to be located at 6.6 Re, i.e. at geosynchronous orbit distance. We obtain a number of useful estimated values. Assuming the smallest ratio $f^2/k$ =1 we obtain minimal value of *Psw* =18.4 nPa. If we take *f*=1 and *k* =0.881 than we obtain *Psw*=20.9 nPa. For the largest theoretical value $f^2/k$=1.69 we have the maximal value of *Psw*=31.1 nPa. Thus, the truth is somewhere between 18.4 nPa and 31 nPa. As one can see from Figure 1 and Table 1 only three models get inside expected theoretical range: PR96, KS98, and DS00 for *By*=20 nT. Rest three models more probably overestimate the SW pressure under positive *Bz*.

For the southward IMF the situation is not so obvious. The main difficult is that nobody knows how the pressure balance changes during reconnection on the MP. It was established empirically by *Rufenach et al.* [1989] that during an erosion of geomagnetic field under strong southward *Bz* the SW dynamic pressure tends to be smaller than under northward *Bz*. The MP models demonstrate decrease of the dynamic pressure under strong negative *Bz* but in different manner (Figure 1). The PR96 and Sh98 models extrapolate from normal SW conditions. The difference between them is explained by the *Bz* saturation effect incorporated in the Sh98 model. In the other three models (KS98, DS00 and Ch02) the *Bz* saturation effect is a result of the GMC data fitting. The difference between predictions of these three models is owing to different GMC data sets and different algorithms used for the data approximation. The main statistical characteristics of the data set used in the KS98 and DS00 are characterized by the distribution functions that are similar to the averaged distributions of the SW parameters indicating that the data used in the models reflect adequately the real SW conditions [*Veselovsky et al.*, 1998; *Dmitriev and Suvorova,* 2000]. Consideration of the SW conditions for magnetopause crossings used in development of the PR96 and Sh98 models show that the most probable value of the SW dynamic pressure is about 1.3 nPa, that is smaller than the most probable value typical for the SW which is varying from 1.5 nPa in solar maximum to ~3 nPa on declining phase of solar activity [*Dmitriev et al.*, 2002a]. It is unclear how the model developed for weaker SW conditions can extrapolate into strongly disturbed conditions in the solar wind. For any empirical models the difference in the modeling data sets necessarily leads to difference in prediction which one can clearly observe in Figure 1. Note again that the curves for KS98, DS00 and Ch02 models are inferred from the data analysis (for both northward and southward IMF) rather than obtained from a successful idea, which is not based on real data, or from model extrapolation of magnetopause crossings from normal conditions. However even the KS98, DS00 and Ch02 models developed using GMC data predict very different solar wind conditions required for GMC.

The contradiction results obtained from statistical comparison of the MP models using parameters PCP and FAR are probably owing to weak effectiveness of the methodic of analysis. Actually the PCP and FAR can not indicate the most realistic MP model which is better describe actual dynamic of the magnetopause under strongly disturbed solar wind conditions. Indeed models with very strong underestimation or overestimation may have the same PCP. So the methodic used in *Shue et al.* [2000b, 2001] gives sometime strange results showing that all three model are good, despite their predictions of the SW pressure (Table 1) are significantly different (up to a factor of two). As it was noted by *Shue et al.* [2000b] in context with SH98 and PR96 the high PCP and FAR indicate that the occurrences of the MP crossings are depended on other necessary conditions. To solve this problem it is very important to obtain empirically significant result. For convincing model comparison the statistical method should be improved and it should combine the PCP with other "orthogonal" statistical parameter such as prediction/observation ratio [*Merka et al.,* 2003] or overestimation/underestimation ratio (OUR) suggested in [*Dmitriev et al.,* 2003a, 2005c].

Finally, two questions are still open. What are actually necessary conditions to observe the GMCs? What is the MP model more adequate describing the observations of the GMCs? An empirically proved answer on the first question is absent now. An answer on the second question, which does not contradict to different investigations, also is not obtained yet, though many comparison analyses of mentioned above models were performed. As we can see the arguments about MP model underestimation/overestimation can not be completely refused or confirmed without a new investigation based on 1-minute dataset with large statistics and taking into account three mostly important factors (2)-(5). In this paper we undertake to answer on the both questions on basis of the largest statistics of the high time resolution data of GMCs from 8 geosynchronous satellites GOES and LANL series accumulated during 1995-2001.

In section 2 we will analyze possible reasons that lead to disagreement of the models with observations. The methodic of data treatment is described in section 3. Section 4 is devoted to derivation of the necessary conditions for GMC. Discussion of results and conclusions are in sections 5 and 6.

## 2. Uncertainties in GMC modeling

### 2.1. Longitudinal Effect

Geomagnetic field generated by internal sources is substantially different from the dipole filed. Figure 2 demonstrates result of the geomagnetic field *H* calculation in equatorial plane (zero geographic latitude) on radial distance $R=6.62\ R_E$ performed by using the pure dipole (dashed curve) and IGRF (solid curve) models for epoch of 1999. Difference between the IGRF model and dipole field varies from $-3$ nT on *Lon*=-45° to 4.5 nT on *Lon*=120°. Hence calculation of the geomagnetic field pressure and, therefore, of the solar wind dynamic pressure required for GMC (see eq. 1.1) gives ~15% difference between the IGRF and dipole models. Furthermore, the geomagnetic field calculated from IGRF model in geosynchronous orbit varies substantially with longitude from $H=102$ nT at *Lon*=0° to $H=113.5$ nT at *Lon*=120°. Due to this variation the geomagnetic field generated by internal sources is different for the geosynchronous satellites located on different longitudes [*Degtyarev et al.*, 1985; *Itoh and Araki*, 1996]. Geographical longitudes corresponding to different geosynchronous satellites are indicated in Figure 2 by vertical lines and listed in Table 2. The magnitudes of the geomagnetic field $H$ modeled using IGRF are also presented in Figure 2 and in Table 2 for different geosynchronous satellites. The abbreviations of the satellite names used in Figure 2 are decoded in the first column of Table 2 in parentheses. To demonstrate the longitudinal effect we consider the geosynchronous satellite LANL-1991 (L1) located on geographic longitude *Lon*=0° where magnitude of the geomagnetic field is $H=102$ nT and LANL-1994 (L4) with location on *Lon*=135° and $H=113$ nT. Because the difference of 11 nT between geomagnetic field magnitudes, the magnetopause crossings of the LANL-1994 orbit requires stronger disturbed solar wind conditions than for the LANL-1991 orbit. The difference in the solar wind dynamic pressure required for GMC observation by LANL-1991 and LANL-1994 is estimated to be about 20%. This difference can be considered as an internal uncertainty in determination of the solar wind conditions for magnetopause crossings observed by different geosynchronous satellites.

## 2.2. Latitudinal and Tilt Angle Effects

The uncertainty growth up when we take into account that the geomagnetic field in geosynchronous orbit can be significantly deviated from the field generated by internal sources due to contribution from the magnetospheric currents such as Chapman-Ferraro, ring current, tail current and field-aligned currents. Modern models of the geomagnetic field use GSM coordinate system for mapping the magnetic field in the outer magnetosphere where geosynchronous orbit is situated. The MP models also operate in GSM coordinate system. During rotation around the Earth the geosynchronous satellites pass different GSM longitudes and latitudes. Figure 3 demonstrates in GSM coordinate system the distributions of probabilities of occurrence in different regions of the dayside sector for different geosynchronous satellites. In Figure 3 one can clearly see that GSM latitude for geosynchronous satellites varies significantly from -26° to 26°. Moreover, in vicinity of the noon (LT~12 h) the occurrence probability at middle latitudes (*Lat*>20°) is much higher (about 100 times) than at the region of "zero" point (*Lat*~0° and LT=12 h). One can expect that most statistics of the geosynchronous magnetopause crossings is actually accumulated on the middle GSM latitudes and longitudes, but only a few GMCs are observed near the magnetopause "zero" point associated traditionally with subsolar and/or nose point.

To estimate the latitudinal effect in the outer magnetosphere magnetic field we have to use a realistic model considering external sources of the geomagnetic field such as T01 [*Tsyganenko,* 2002] presented also on web-site http://nssdc.gsfc.nasa.gov/space/model/magnetos/data-based/whatnew6.html. Figure 4 demonstrates the total geomagnetic field distribution calculated from T01 model in GSM coordinates for the MP dayside sector under different solar wind dynamic pressures: *Pd*=4 nPa (a) and *Pd*=9 nPa (b). The other model parameters are accepted as *Bz*=0 nT, *By*=5 nT, *Dst*=0 nT and the dipole tilt angle *PS*~0°. For moderate solar wind dynamic pressure (Figure 4 a) the geomagnetic field distribution is practically symmetric and gradually decreases from the "zero" point with rate of about 2-4 nT per 10°. We have to indicate that the decrease of the geomagnetic field is not monotonous, but it has a local minimum at latitudes of about 20°. For strong solar wind dynamic pressure (Figure 4 b) the geomagnetic field distribution has substantial dawn-dusk versus south-north asymmetry. The field magnitude in the noon region decreases on ~30 nT (from 150 nT to 125 nT) when the GSM latitude is changed from 0° to

-20° while longitudinal gradient of the geomagnetic field is about three times smaller (only ~10 nT per 20°). Hence under strongly disturbed solar wind conditions the magnetopause formation on middle latitudes should be differ than in the GSM equator and the difference can achieve tens percents.

The latitudinal effect is mostly controlled by contribution of the mentioned above magnetospheric currents influencing significantly on the geomagnetic filed in geosynchronous orbit. This influence can change substantially the conditions for the pressure balance on the magnetopause. However the evaluation of this effect is very complicated. For example during severe geomagnetic storms when *Dst*~–100 nT (Figure 4 c) the T01 model shows the geomagnetic field asymmetry but with reverse gradients such that the geomagnetic field increases with latitude faster than with longitude.

The situation becomes more complicated if we take into account variations of the tilt angle *PS*. This is angle between the axis of Earth's centered dipole moment and Z-GSM axis (PS<0 in winter and PS>0 in summer). Figure 4 d demonstrates the geomagnetic field distribution calculated for *PS*~20° and *Pd*=4 nPa. The distribution has south-north asymmetry such that at GSM latitude about -15° the geomagnetic field has a local maximum with *H*~130 nT associated with highest contribution from Chapman Ferraro current in the mostly compressed region of the magnetosphere. This effect leads to latitudinal shift of the MP nose point from the "zero" point in GSM [*Petrinec and Russell,* 1995]. We determine here the MP nose as a point where the normal to the magnetopause is aligned the solar wind flow directed along X-axis. In the case presented in Figure 4 d we can expect that the MP nose point shifts toward the region with mostly compressed geomagnetic field which is located on the southern latitudes. Above presented examples clearly demonstrate that the nose point continuously moves in a vicinity of the "zero" point due to tilt angle variations. Unfortunately it is practically impossible to account strictly all above mentioned effects for the GMCs because most existing models of the geomagnetic field are limited by quiet and moderately disturbed conditions and, thus, can not be used for GMCs caused by strong geomagnetic and solar wind disturbances.

# 3. Data Treatment

## 3.1. Driving Parameters and Coordinate System

For identification of the GMCs we use high-resolution (~1 minute) ISTP data (http://cdaweb.gsfc.nasa.gov/cdaweb/istp_public/) of the geosynchronous satellites (GOES and LANL) and upstream monitors Geotail, Wind, and ACE for time period from 1995 to 2001.

The GMCs are observed under strongly disturbed solar wind conditions which are often accompanied by large non-radial components of the solar wind velocity (more than 100 km/s), enhanced He/H ratio (up to few of tens percents), IMF strength and solar wind plasma temperature. Therefore, in the study of geosynchronous magnetopause crossings we have to take into account non-radial propagation of the solar wind and, thus, to use aberrated GSM (aGSM) coordinate system. We have also to consider a total solar wind pressure $Psw$ (SW pressure). In the first approach, the SW pressure in vicinity of the MP nose point can be accepted as a sum of the solar wind dynamic $P_d$, thermal $P_t$ and magnetic $P_m$ pressures:

$$Psw = Pd + Pt + Pm \quad (3.1)$$

$$P_d = 1.672 \cdot 10^{-6} \cdot D \cdot V^2 \quad (3.2)$$

$$P_t = 1.6 \; 10^{-4} \; D \; T \quad (3.3)$$

$$P_m = 3.98 \; 10^{-4} \; B^2 \quad (3.4)$$

Here $P_d$, $P_t$ and $P_m$ are calculated in nPa; $T$ and $B$ are, respectively, solar wind temperature (in eV) and IMF strength (in nT). The solar wind density $D$ (cm$^{-3}$) is calculated using helium contribution:

$$D = n(1 + 4HeP) \quad (3.5)$$

where *n* is proton concentration in the solar wind (in cm$^{-3}$) and *HeP* is helium/proton ratio. The helium/proton ratio can be obtained from the ACE data or can be applied to be 4% if the ACE data is unavailable.

In transformation of the GSM into aGSM we take into account two effects: 1. Earth's orbital rotation around the Sun with speed $V_E$=30 km/s in the GSE coordinate system; 2. non-radial solar wind propagation when Y and/or Z GSE components of the solar wind velocity may increase up to 100 km/s or even more. The GSM coordinate system is aberrated such that X-axis of the aGMS becomes aligned solar wind flow. The transformation is based on two consequent rotations of the GSE coordinate system and than standard conversion of the aberrated GSE (aGSE) into aberrated GSM (aGSM). The transformation from the GSE to aGSE consists of two consequent steps:

1. Rotation about Z-GSE axis on the angle $\delta_Y$:

$$\delta_Y = \tan^{-1}(\frac{V_y + 30}{|V_x|}) \quad (3.6)$$

2. Rotation about aberrated Y-axis on the angle $\delta_Z$:

$$\delta_Z = \tan^{-1}(\frac{V_z}{\sqrt{V_x^2 + (V_y + 30)^2}}) \quad (3.7)$$

Here $V_x$, $V_y$, $V_z$ are, respectively, X, Y, Z components of the solar wind velocity in GSE. The transformation into aGSE described by equations (3.6) and (3.7) and following conversion of aGSE to aGSM are applied for both the satellite coordinates and for the IMF components. The solar wind velocity in aGSM has only one nonzero X-component. Moreover, *Bz* in aGSM can be different from the *Bz* in GSM. In the aGSM *Bz* is perpendicular to the solar wind vector and, hence, it should be tangential to the magnetopause surface in the nose point.

## 3.2. Method for Solar Wind Timing

For 1-min time resolution we suggest practically instant (1-2 min) response of the magnetopause on changing of the interplanetary conditions. Such suggestion is supported by the direct comparison of the magnetopause dynamics with corresponding variations of the *Psw* and *Bz* in the interplanetary medium. Valid determination of the corresponding upstream solar wind conditions (i.e. timing) is based on time delay for direct propagation of entire solar wind structure observed by an upstream solar wind monitor, and on additional time shift associated with tilted interplanetary front. The final timing is determined using two independent criteria. The one general criterion is based on dependence of the *Dst* index on a square root of the SW dynamic pressure, caused by pressure-associated changing of the Chapman-Ferraro current on the magnetopause [e.g., *Burton,* 1975; *Russell et al.*, 1994a,b]. We calculate the cross-correlation between the solar wind pressure and 1-min resolution *Dst* (SYM) index obtained from http://swdcwww.kugi.kyoto-u.ac.jp/aeasy/index.html. The SW pressure is compared with *Dst*-deduced pressure $P_{Dst}$:

$$P_{Dst} = \left(\frac{Dst - a}{b}\right)^2 \qquad (3.8)$$

where parameters *a* and *b* are obtained as regression coefficients for equation:

$$Dst_i = a + b\sqrt{Psw_i} \qquad (3.9)$$

where $Dst_i$ and $Psw_i$ are magnitudes of the 1-min *Dst* variation and solar wind pressure at a time moment $t_i$. The regression is calculated for the entire considered interval which duration is usually about one or a few hours. The best correlation between SW pressure and *Dst* indicates in most cases the best timing as well as the best choice of the upstream monitor. The efficiency and objectivity of this method are obvious.

The second, subsidiary, criterion for the upstream solar wind timing for the GOES satellites is co-variation of the *Bz*, *By*, and *Bx* components of the magnetic field measured by GOES during magnetosheath intervals and IMF components measured by an upstream solar wind monitor. For the LANL satellites the similar independent criterion is co-variation of the ion density in the magnetosheath with the solar wind pressure. The accuracy of such method based on ~1-minute time resolution of the experimental data is estimated from 1 to 2 minutes. The accuracy can be affected by non-instant magnetopause response, continuous changing of the solar wind front tilting [*Collier et al.*, 1998], fast variations of the solar wind plasma and IMF properties [*Richardson and Paularena*, 2001; *Werner et al.*, 2002] and evolution of the SW irregularities propagating through the interplanetary medium and the magnetosheath.

A problem of choice of the monitor for better forecasting the near-Earth environment was formulated in [*Song et al.*, 2001]. We propose a good way to solve the problem of properly choosing the SW monitor for any case event in the magnetosphere. Our 'diagnostics' method based on two important factors considered together: the satellite location and the relation of the *Dst* with the *Psw* measured by the satellite. Method is very useful when there are a few SW monitors. Fortunately, we often have two or three SW monitors with very different locations. It is well known that ACE locates always near the liberation point L1, Geotail is on near-Earth orbit, and Wind position varies in wide spatial range with apogee of about 220 Re. Initially, we visually track the spatial and time variation of the SW and IMF parameters by comparing the data of one upstream monitor (for instance, ACE) with other. There are three situations: quantitative coincidence of the data of the different satellites, coincidence of the variations with offset between their average values, and absolutely different data. Obviously, the choice of the best monitor (or which satellite location better forecasts) practically does not matter for the first, the simplest situation. Second and third situations can be successfully resolved only if we use a diagnostics method based on (*Dst*,*Psw*) relation. Concerning the satellite locations, that are important too, we prefer do not use a monitor located too far from the flank of the magnetosphere (more than 70 Re) [*Richardson and Paularena*, 2001; *Riazantseva et al.,* 2002] and too close to the bow shock (less than 30 Re) [*Prech et al.,* 2005]. We must emphasize that studying case by case we find for most of them a very good correlation between the *Dst* variation and the data from the monitor located at L1. This is in good agreement with study of the interplanetary and magnetosheath conditions by [*Song et al.*, 2001]. The efficiency and objectivity of the proposed method we will demonstrate in the next sections.

### 3.3. GMC Identification from GOES Magnetic Field Data

The GMC is identified using GOES magnetometer data when one of two requirements is satisfied: 1) the magnetic field measured by GOES is significantly deviated from the geomagnetic field, 2) the components of the magnetic field measured by GOES correlate with the IMF components measured by an upstream monitor. Both rules help us to find MP crossings and magnetosheath intervals even under northward *Bz*, which are more difficult to identify. In Plot 1 we demonstrate the GMC identification using GOES 8 magnetic measurements on September 18, 2000. The ACE satellite is selected as upstream monitor. Two top panels on Plot 1 present predictions of the magnetopause models by Ch02 and Sh98 (solid and dotted curves respectively) and by KS98 and DS00 (solid and dotted curves respectively). The third panel (from top to bottom) shows *Hp* component and total magnetic field (nT) measured by GOES (solid and dotted curves respectively). Solar wind pressure and *Dst*-deduced pressure $P_{Dst}$ are presented by solid and dotted curves respectively on the fourth panel. Three next panels show *Bz*, *By*, and *Bx* magnetic field components in aGSM measured by ACE and by GOES (solid and dotted curves respectively). Aberration angles $\delta_Y$ and $\delta_Z$ (in degrees) are indicated by solid and dotted curves respectively on the eighth panel. The latitude (in degrees) and local time (in hours) of GOES satellite in aGSM coordinate system are presented on two bottom panels respectively. Vertical dotted and dashed lines indicate magnetosheath entrance and exit respectively. The *Bz* and *By* components measured by GOES are divided on 10 and *Bx* component is divided on 5.

The GMC is observed during 5 min within long-lasting (about one hour) solar wind pressure enhancement (up to 20 nPa) intersecting the Earth form 1445 UT to 1560 UT ($LT$=10~11 h). The $Bz$ rotates from strong southward (-10 nT) to strong northward (10 nT). The pressure enhancement reveals itself in positive $Dst$ variation as well as in significant increase of the geosynchronous magnetic field, which arises up to 250 nT. Using the method of cross-correlation between $Psw$ and $P_{Dst}$, we obtain 28-min time shift for the ACE data to achieve the best coincidence between the variations of the solar wind pressure and geomagnetic field. Note that our time shift is less on 4 min than the average time calculated from SW direct propagation. The fourth panel demonstrates very good coincidence of $P_{Dst}$ with $Psw$ and very high correlation coefficient between the SW pressure and $Dst$ variation ($r$=0.97) despite long-lasting interval (~30 min) of negative $Bz$. A large decrease of the $Hp$ and $Bt$ measured by GOES-8 indicates that the magnetopause enters inside the geosynchronous orbit at 1509 UT and comes back at 1514 UT. The GMC is caused by short enhancement of the SW pressure up to 18 nPa. The small southward IMF is also observed during this interval but it has secondary role in providing the GMC. Indeed, despite increase of the negative $Bz$ up to about –7 nT from 1513 UT to 1516 UT, the magnetopause exit is observed at 1514 UT and caused mainly by decrease of the SW pressure.

Because the negative $Bz$ during the magnetosheath interval is small, the $Hp$ and $Bz$ components measured on GOES have negative values during only one-two minutes near 1512 UT. The difference in the behavior of relatively small $Bz$ component measured by GOES-8 and ACE can be attributed to magnetic field draping in the magnetosheath [*Yang et al.,* 2002]. However, during the magnetosheath interval we observe very high co-variations of the $Bx$ and $By$ components detected by GOES-8 in the magnetosheath and by ACE in the interplanetary medium. This co-variation also supports our suggestion that the GOES-8 is located in the magnetosheath, and that the time shift for the ACE data is determined correctly.

It is interesting to note that all the four MP models can not predict this GMC. On the other hand the Sh98 and KS98 predict false magnetosheath interval from 1447 UT to 1452 UT which could be caused by strong increase of the SW pressure (~18 nPa) and large negative $Bz$~-10 nT. The absent of the GMC at 1447 UT and its presence at 1509 can be attributed to half-hour difference in their local time: LT~10 h for the false GMC and LT~10.5 h for the observed one.

### 3.4. GMC Identification from LANL Plasma Data

Unfortunately LANL data from ISTP database do not contain information about MPA spectral measurements. Therefore the methods for GMC identification proposed by *McComas et al.* [1994] as well as by *Shue et al.* [2001] can not be applied for these data. In this situation we suggest another method for GMC identification using ISTP data on energetic ion and electron fluxes calculated from MPA spectral measurements. We introduce a ratio:

$$RI = D_{hip}/T_{hip} \quad (3.10)$$

Where $D_{hip}$ and $T_{hip}$ are respectively density (cm$^{-3}$) and temperature (keV) of the high energy ions (E=130eV/q - 45keV/q). The $T_{hip}$ is calculated from its perpendicular ($T_{per}$) and parallel ($T_{par}$) components:

$$T_{hip} = (2 \cdot T_{per} + T_{par})/3 \quad (3.11)$$

In the dayside magnetosphere where the average high-energy ion density $D_{hip}$~1 cm$^{-3}$ and temperature $T_{hip}$~10 keV, the ratio *RI* is about 0.1. In the magnetosheath the ion temperature is ~100 times smaller and the density is ~100 times larger, or by other words, $T_{hip}$ is about a few tenth of keV and $D_{hip}$ is about a few of tens particles in cm$^3$. Hence we can accept the *RI* threshold value for the magnetosheath as *RI*>30. The intermediate region of *RI* between several tenths and a few tens corresponds to the low latitude boundary layer (LLBL).

An additional indicator of the magnetosheath interval is significant decrease of temperature $T_e$ and increase of density for electrons in energy range from 30 eV to 45keV. The average temperature $T_e$ in the magnetosphere is about a few of keV. In the magnetosheath the electron temperature drops down to 0.1 keV or even less. The electron density in the magnetosphere is about $D_e$=1 cm$^{-3}$ and in the magnetosheath the density increases significantly up to few of tens electrons per cm$^3$. So we can introduce additional (secondary) indicator for the magnetosheath conditions:

$$RE = D_e/T_e \qquad (3.12)$$

The *RE* threshold for the magnetosheath is accepted to be *RE*>100. In spite of that clear terminology, it is not always easy to distinguish LLBL and magnetosheath regions on the basis of their plasma signature alone. However, we have to take into account that even if the satellites are in the LLBL and have not actually crossed the magnetopause, the magnetopause must nonetheless be quite close because we are seeing magnetosheath plasma that has entered the magnetosphere, presumably by dayside reconnection (i.e., the satellite is on field lines that are open or at least have been open quite recently).

An example of the GMC identification using LANL data on May 24, 2000 is presented on Plot 2 including following panels (from top to bottom): prediction of the MP location by Ch02 and Sh98 models (solid and dotted curves respectively); predictions of the MP location by KS98 and DS00 models (solid and dotted curves respectively); ratios *RI* and *RE* (solid and dotted curves respectively); SW pressure and $P_{Dst}$ (solid and dotted curves respectively); IMF *Bz* and *By* components (solid and dotted curves respectively) in aGSM; aberration angles $\delta_Y$ and $\delta_Z$ in degrees (solid and dotted curves respectively); aberrated geographic latitude and geomagnetic latitude (in degrees) of the LANL satellite; geographic local time (in hours) of the LANL satellite corrected on the aberration. Vertical dotted and dashed lines indicate magnetosheath entrance and exit respectively. Horizontal dashed lines on the third panel indicate the thresholds for *RI* and *RE*. It is important to note that geomagnetic latitude is associated with the Earth dipole coordinate system and it is absolutely different from the GSM latitude.

Unfortunately ISTP data for LANL satellite do not contain information about the LANL location in GSE/GSM coordinate system. The LANL location in aGSM is calculated separately using GEOPACK program package (http://nssdc.gsfc.nasa.gov/space/model/magnetos/data-based/geopack.html).

Long-lasting magnetosheath interval (~2 hours) is observed by LANL-1994 from 0156 UT to 0415 UT on May 24, 2000 when the ratios *RI* and *RE* exceed their thresholds significantly. During this period LANL-1994 moves from LT~9 h to LT~11.5 h. The Wind satellite used as upstream solar wind monitor is located close to the Earth (45, 5, -3) Re GSM. The method of cross-correlation between the *Psw* and *Dst* gives the time delay on the solar wind propagation of about 6 min, which coincides with average time of the SW direct propagation. However the best correlation coefficient is small ($r=0.37$). The weak correlation between the SW pressure and *Dst* variation is explained by strong contribution from the ring and tail currents intensified due to continuous large southward IMF observed from 0120 UT to 0340 UT. Manifestation of the influence of the currents is the gradual decrease of the *Dst*. Nevertheless, some fine structures in the *Dst* variation coincident with large variations of the SW pressure during intervals at 0228-0233 UT, 0345-0416 UT and 0432-0438 UT. Moreover both the *RI* and *RE* correlate very well with the SW pressure. These co-variations support the correct timing for the Wind data as well as correct choice of the upstream monitor.

The magnetosheath interval is accompanied by strong magnetosphere compression and erosion caused, respectively, by high SW pressure (up to 36 nPa) and large negative *Bz* (up to -32 nT). However the GMC at 0156 UT is mostly caused by LANL motion from early morning toward the noon. Indeed before the GMC the solar wind conditions were strongly disturbed such that some magnetopause models (Sh98, KS98, DS00) predict false GMCs. But such SW conditions are still insufficient to produce GMC near the dawn flank. The magnetosheath exit at 0415 UT is caused by sharp decrease of the SW dynamic pressure. The other magnetosheath interval observed by LANL-1994 from 0432 UT to 0433 UT on May 24, 2000 is very short. This interval is caused by short increase of the SW pressure accompanied by negative IMF *Bz*.

It is interesting to note that predictions of the MP models are often incorrect in this case event. A part of the false predictions is corrected when we take into account the He contribution measured by ACE. For this event the He contribution is more than 10 %. The first false magnetosheath exit at ~0207 UT is predicted by the Sh98, KS98, Ch02 models, while the DS00 model provide correct prediction. From ~0340 UT to 0415 UT all the models overestimate the MP distance and can not predict magnetosheath interval. From 0420 UT to 0432 UT we observe opposite situation when Sh98, KS98, Ch02 models substantially underestimate the distance to the MP and predict false magnetosheath interval. However the DS00 model has a correct tendency to predict the MP location outside the geosynchronous orbit. This example demonstrates one important advantage of the DS00 model, which introduces a dependence of the MP distance on the IMF *By* absolute value. Indeed, during the time interval 0420-0432 UT when the IMF turns southward, the MP models predict significant decrease of the MP distance. At the same time the IMF *By* absolute value decreases from about −20 nT to about −5 nT. Hence, despite large negative IMF *Bz* (up to −20 nT), the DS00 model predicts a small change of the MP distance anti-correlating with *By* variation. We have to note that in this comparison we calculate the MP model predictions without information about actual location of the LANL-1994 satellite. Consideration of the aGSM latitude for the geosynchronous satellite can significantly modify the model predictions especially in the noon region.

### 3.5. Surface waves

During GMC identification we find several problematic cases associated with high amplitude wave activity on the magnetopause. In such GMC events there is no direct relationship between variations of the interplanetary conditions and magnetopause dynamics. One of such so-called "wave events" on October 21, 2001 is presented on Plot 3. We consider data from GOES-8 (*Lon*~75° deg), GOES-10 (*Lon*~135°) and LANL-1991 (*Lon*~165°). The GOES-8 (Plot 3 a) is the first satellite observing the dayside magnetopause, then 4 hours later the GOES-10 (Plot 3 b,c) is going and after 2 hours the LANL-1991 (Plot 3 d) closes the observations. Description of the panels on Plot 3 is the same as on Plot 1 and Plot 2 for GOES and LANL satellites respectively.

The GOES-8 (Plot 3a) observes the wave structure since ~19 UT (LT~15h) as multiply GMCs. At ~1905UT and 1942-1947UT we can see large amplitude wave-like fluctuations of the GOES-8 magnetic field components. Remember that *Bz* and *By* components are re-scaled on 10 and *Bx* is re-scaled on 5. At the same time (from ~19 UT to 1930 UT) the GOES-10 (Plot 3b) observes a wave structure in the prenoon sector at LT~10.5 h. The main feature of this wave structure is anticorrelation between *Bz* and *Bx*. Comparison between the GOES-8 (postnoon sector) and GOES-10 (prenoon sector) shows that the component *By* correlates with *Bz* in the prenoon and anti-correlates in the postnoon sector. Because the *By* component change the sign from positive in prenoon to negative in postnoon, the *By* correlation/anticorrelation with *Bz* is owing to redistribution of the magnetic field between *Y* and *Z* components. The variation of the *Bx* is altering in the sign. It means that the GOES magnetic field vector rotates around an axis laying practically in *YZ* plane. Undoubtedly further detail study of the wave structures is required but this is beyond our current work.

At ~2315 UT (LT~15h) the GOES-10 (Plot 3c) comes in the wave structure which was observed before by the GOES-8 and seems it until midnight. Large amplitude fluctuation of the *RI* and *RE* on the LANL-1991 during time interval 0020-0120 UT on October 22 (postnoon LT~13.5-14.5h) have also wave-like structure where *RI* correlates with *RE*. At some moments the fluctuations become so large that the LANL-1991 observes multiply GMCs at 0050-0110UT on October 22. The partial densities of ions above (hp) and below (lp) 100 eV, as well as the derived flow velocities for those two populations (Vlp and Vhp) are shown on Plot 4 for considering time interval from 0000 to 0130 UT on 22 October 2001. The velocities are computed in the spacecraft frame, with +X in the northward direction, +Y in the eastward direction, and +Z radially inward. Plot 4 illustrates clearly that there were flow oscillations associated with the in/out variability of the magnetopause location. Those velocity oscillations appear to persist for a while after the satellite re-enters the magnetosphere at 0107 UT, for maybe an hour.

Therefore we find that the wave structure is observed by the three satellites at practically the same LT-location during more than 6 hours from ~19 UT on October 21 to ~1 UT on October 22. Furthermore, this structure is observed simultaneously both in prenoon and postnoon sectors. We can interpret this structure as long-living standing waves on the magnetopause associated with strong magnetosphere compression caused by large solar wind dynamic pressure.

The "negative" property of the wave structures is that the magnetopause dynamics (variations of the size and shape) is not controlled directly by the SW pressure and IMF $Bz$. Hence we have to exclude the wave events from our consideration of the direct relationship between the MP size and interplanetary conditions for GMCs. We find 7 "wave" events and exclude them from our study of GMC events.

## 4. Solar Wind Conditions for the GMCs

Using described above methods we identify for the GOES and LANL satellites the magnetosheath entrances and exits together with magnetosheath intervals (so called GMC intervals) and determine for them the upstream solar wind conditions. Common statistics of considered case events and identified GMC intervals is presented in Table 3. The sum in the third column consists of numbers of magnetosheath entrances and exits that can be different due to the data gaps. One can see that the number of GMCs and total duration of the magnetosheath intervals observed by GOES and LANL satellites are comparable. Therefore our data set can be considered as homogenous in the sense of the experimental methodic for the GMC identification.

Scatter plot of the selected GMCs in aGSM coordinates latitude versus local time is presented on Figure 5. Triangles indicate measurements during the magnetosheath intervals. Black and gray crosses indicate magnetosheath entrances for GOES and LANL satellites respectively. After Figure 3 it is not surprising that the GMC location varies in so wide range of aGSM latitudes from −28° to 28°. The distribution of the GMCs is not uniform and only a few of crossings are observed in vicinity of the "zero" point. The GMCs are located mostly at conic angles of about $\theta=10°\sim20°$. The conic angle is calculated from the X-aGSM axis. Such location of the GMCs can be explained by superposition of two opposite effects. On the one hand, the occurrence probability for the geosynchronous satellite location increases faster with latitude and longitude (Figure 3). On the other hand, the probability to observe the MP on larger latitudes and longitudes should decrease because of blunted MP shape. There is no substantial dawn-dusk asymmetry of the dayside magnetopause crossings in the geosynchronous orbit for entire data set of the GMCs.

Scatter plot of the GMCs in space of the solar wind parameters SW pressure versus aGSM $Bz$ is presented in Figure 6. The solar wind conditions vary in very wide range of the $Psw$ (from ~4 nPa to 100 nPa) and aGSM $Bz$ (from –40 nT to 40 nT). One can see that the solar wind conditions requiring for GMCs are enveloped by quite sharp boundary indicated by thick solid line, under which GMCs do not occurred excepting a few magnetosheath points. The magnetosheath points situated under the boundary can be attributed either to influence of "unknown" effects and, consequently, parameters controlling the MP dynamics or to experimental data "noise" associated with evolution of the solar wind conditions during propagation from the upstream monitor to the Earth. Numerical determination of the envelope boundary requires taking into account these uncertainties and selection only meaningful measurements.

Let us describe the method of the boundary derivation. Firstly, we can distinguish two strait horizontal branches of the boundary where $Bz$ does not influence on the magnetopause location. The first branch is located in the range of $Psw\sim20$ nPa for large positive $Bz>20$ nT and associated with a pressure balance regime. The second branch approaches the range of $Psw\sim5$ nPa under very strong negative $Bz$ and it is associated with a so-called "regime of $Bz$ influence saturation". Secondly, for the moderate positive and negative IMF $Bz$ we can indicate the intermediate region of the boundary where the minimal SW pressure required for GMC decreases gradually in response to decreasing $Bz$. The horizontal branches as well as intermediate region can be approximated by a hyperbolic tangent function [*Dmitriev and Suvorova,* 2000]:

$$Psw(Bz) = P_0 - \frac{dP}{1+\exp\{\chi(Bz+Bz_0)\}} \qquad (4.1)$$

The variables $P_0$ and $dP$ may be estimated as asymptotes of the function $Psw(Bz)$ when $Bz \rightarrow -\infty$ and $Bz \rightarrow +\infty$, respectively. The coefficients $\chi$ and $Bz_0$ are calculated by a simple approximation of the points located in close vicinity of the boundary.

To estimate the asymptotes and select the points for approximation of the intermediate region we analyze two-dimensional distribution of occurrence number for the GMC intervals (Figure 7). The space of the solar wind parameters $Psw$ versus $Bz$ is split on 20x20 bins with width $\Delta Bz=3$ nT and height increasing logarithmically with $Psw$. For each bin the number of magnetosheath points and magnetosheath entrances weighted on 3 are calculated. We artificially increase the statistics for the magnetosheath entrances because during selection of the GMC intervals the entrance to the magnetosheath is considered as a 'bench mark' point. The occurrence number in different bins is indicated in pseudo-logarithmic gray scale and varies from <10 (white bins) to >300 (black bins). To select the meaningful events we accept the occurrence number 10 as a lower threshold for the meaningful statistics that is less than 3% of the maximal occurrence number in the bin (~400). As we can see from Figure 6 such suggestion provides a reasonable envelope boundary consisting on meaningful edge bins that are corresponded to minimal SW pressures required for GMCs. On this boundary for each given $Bz$ the occurrence number decreases sharply (several times) when the SW dynamic pressure goes down from edge bins (gray color) with occurrence number >10 to white bins with "noise" data where occurrence number <10 (Figure 7).

For the asymptotes we accept the minimal pressures for magnetosheath entrances under strong positive (>15 nT) and strong negative (<-15 nT) IMF $Bz$. There are $P_0$=21 nPa for $Bz$~17 nT and $P_1$=4.8 nPa for $Bz$~-29 nT and, therefore, $dP=P_0-P_1$=16.2 nPa. Using the magnetosheath points and entrances from the edge bins that occurred under SW pressures $P_1$ and above $P_0$, we obtain a data subset for approximation in the intermediate region. Taking into account that $dP=P_0-P_1$, we can rewrite the equation (4.1) as:

$$F(Psw) \equiv \ln\left(\frac{Psw - P_1}{P_0 - Psw}\right) = \chi(Bz + Bz_0) \qquad (4.2)$$

Assigning the empirical asymptotic values to $P_0$ and $P_1$, the coefficients $\chi$ and $Bz_0$ in the equation (4.2) are calculated from approximation of the $F(Psw)$ by a linear function of the $Bz$. Figure 8 shows the points from the data subset and their linear approximation (solid line) plotted in the coordinates $F(Psw)$ versus $Bz$. One can see that the points from the edge bins are located very close to the approximation line (RMSD=0.37) in very large dynamic range of the $Bz$ varying from –20 nT to 17 nT and of $F(Psw)$ varying from –3.9 to 3.7 (strictly indicating that our approach is reasonable. The approximation gives us the following values of the coefficient from (4.1): $\chi$=0.2 and $Bz_0$=0.6. Thereby, we derive a numerical expression for the boundary enveloping minimal solar wind conditions required for the geosynchronous magnetopause crossings:

$$Psw = 21 - \frac{16.2}{1 + \exp\{0.2(Bz + 0.6)\}} \qquad (4.3)$$

We have to note here that change of the asymptotes $P_0$ (say 20 or 22) and $P_1$ (say 4.7 or 4.9) lead to substantial increase of the RMSD in the approximation of the $F(Psw)$ that convince us in correct choice of the asymptotes.

Figure 9 presents the magnetosheath points (triangles) and entrances (crosses) from 20% vicinity of the envelope boundary in aGSM coordinates latitude versus local time. We choose 20% corridor owing to internal uncertainty of the GMC method discussed in section 2.2. The spatial distribution of the GMC intervals has three features. Firstly, Figure 9 clearly shows a dawn-dusk asymmetry in the GMCs, i.e there are 19 crossings in the prenoon sector, but only 7 crossings in the postnoon sector. The magnetosheath points are also asymmetrically distributed. Moreover one can see that GMC intervals spread inside 4-5 hours pre noon and only inside 2 hours after noon, i.e. morning hours are more favorable for GMC intervals than evening hours. The same feature (6 prenoon hours and 3 postnoon hours) was demonstrated in previous studies [*Kuznetsov and Suvorova*, 1997; *Itoh and Araki*, 1996]. Secondly, as one can expect from the orbital covering (Figure 3) the longitudinal interval for GMCs is larger (105°) than latitudinal interval (40°). Thirdly, due to the orbital covering the number of points (6 GMCs) in vicinity of the "zero" point confined in 20°×30° sector is much smaller than in the conic angles more than 10° (20 GMCs at *LT*<11 h or *LT*>13 h). Such spottiness in the spatial distribution of the GMCs corresponding to minimal solar wind conditions indicates on the MP asymmetry under strongly disturbed solar wind conditions.

The list of the magnetosheath entrances associated with envelope boundary is shown in Table 4 with indicating date and time of the GMC observation, geosynchronous satellite observing the GMC (GS), corresponding upstream monitor (UM), aGSM location of the GMC (LT and latitude), magnetosphere tilt angle PS, solar wind conditions (*Psw*, *Bz* and *By* in aGSM) and 1-min *Dst* variation. The boundary enveloping the necessary SW conditions for GMC is supplied by very inhomogeneous data on 26 magnetosheath entrances observed in 17 events by different GOES and LANL satellites. The SW conditions for the GMCs are measured by different upstream monitors. Therefore, we can affirm that the boundary is not originated from systematical error in the upstream SW data or from methodic of the GMC identification. By other words the boundary as well as associated phenomena can not be an artifact.

Figure 10 shows a scatter plot of the magnetosheath entrances form Table 4 in aGSM coordinates LT versus IMF *Bz*. One can see that the local time of GMCs has a tendency to decrease with the IMF *Bz*. Fitting of this tendency by a linear function gives us an approximate relationship:

$$LT(h) = 11.2 + 0.037 * Bz(nT) \quad (4.4)$$

This relationship shows that the dawn-dusk asymmetry of the magnetopause increases in response to decreasing *Bz*. For large positive *Bz* (~20 nT) the location of the GMCs approaches noon. When the *Bz* is large negative (say –20 nT) the GMCs shift toward the dawn at *LT*~10.5 h. We have to note that the expression presented by equation (4.4) is obtained from fitting of widely spread set of points and, thus, it can be considered only as preliminary and approximate dependence of the MP dawn-dusk asymmetry on the IMF *Bz*. Apparently the spreading of the GMCs observed in Figure 10 is owing to contribution of other effects influencing on the dayside MP shape.

Figure 11 illustrates a dependence on dipole tilt angle *PS* for the latitude $\lambda$ of the GMCs from Table 4 and magnetosheath points from 20% vicinity of the enveloping boundary. The dependence can be very well approximated by linear function:

$$\lambda = -1.7 - 0.52 * PS \quad (4.5)$$

where latitude $\lambda$ and tilt angle *PS* are expressed in degrees. As one can see the tilt angle effect influences on the GMC location significantly and causes its latitudinal shift equal to about half of the *PS* angle. Hence even for given SW conditions the GMC location varies with the tilt angle *PS*. Neglecting this effect leads inevitably to errors in a model prediction of the upstream solar wind conditions for the GMCs.

## 5. Discussion

The boundary presented on Figure 6 and described by Equation (4.3) envelopes the minimal conditions required for GMC in the SW pressure and IMF $Bz$, as the main parameters controlling the magnetopause. These necessary conditions cause the MP crossing of the geosynchronous orbit in the point where the magnetopause mostly approaches the Earth. Hereafter we call this point the MP "perigee" point. As we have shown in the previous section the location of the "perigee" point varies widely in latitudes and longitudes.

The latitudinal variations of the "perigee" point can be explained by the dipole tilt angle effect. As shown in [*Petrinec* and *Russell,* 1995; *Sotirelis and Meng,* 1999; *Boardsen et al.*, 2000] the orientation of the dayside magnetopause relatively to the GSM equatorial plane significantly varies with the dipole tilt angle such that the northern cusp shifts toward the equator when the tilt angle is positive. Consequently the other points of the MP should shift toward south and, thus, their latitudes should decrease in response to increasing tilt angle. Figure 11 demonstrate exactly the same dependence of the GMC latitude on the *PS*. Hence the dependence (Equation 4.4) can be interpreted as a dependence of the average aGSM latitude of the MP "perigee" point on the dipole tilt angle for strongly disturbed SW conditions.

Character of the longitudinal variations of the MP "perigee" point (Figure 9) indicates to dawn-dusk magnetopause asymmetry. The problem of the dayside magnetopause dawn-dusk asymmetry for the geosynchronous magnetopause crossings was intensively discussed in the literature. Some studies [*Wrenn et al.*, 1981; *Rufenach et al.*, 1989; *McComas et al.,* 1993; *Itoh and Araki,* 1996; *Kuznetsov and Suvorova*, 1997,1998b; *Dmitriev et al.*, 2001b; 2002b] indicate that GMCs are more often observed on the dawn side and therefore the magnetopause (MP) has dawn-dusk asymmetry during the GMCs. *McComas et al.* [1993] discussed interesting observational effects in geosynchronous orbit such as plasmaspheric population and lobe encounters at different local times that support the suggestion on the asymmetrical shape of the magnetosphere [*Rufenach et al.*, 1989]. For instance, they found the strong preference for observing the rare lobe events on the postmidnight in comparison with premidnigth side of the magnetosphere. Moreover, the lobe intervals generally associated with unusual upstream conditions have clear tendency to occur within a day or two of GMCs. But later *McComas et al.*, [1994] present evidences against the asymmetry and show that the relatively small shift of about half an hour in LT toward the morning for most GMC observations may be simply explained by the effects of solar wind aberration. Also note, however, that their conclusion based on only a few events presented as examples of the symmetric magnetopause shape.

In the previous section we have found that the GMCs from vicinity of the envelope boundary demonstrate a tendency to shift toward morning hours under strong negative $Bz$ (Figure 10). The same tendency is predicted by the KS98 and DS00 models that suggest a large-scale MP asymmetry on the dayside. In the KS98 model the entire magnetopause shifts toward the dusk for southward IMF such that the morning and prenoon sectors approaches the Earth. In the DS00 model the dayside magnetopause shape changes nonlinearly (see Figure 4 in the original paper) and the prenoon sector is usually closer to the Earth than postnoon sector of the magnetopause especially for the strong negative IMF $Bz$. Hence we can suggest that the MP "perigee" point longitude increase toward dawn in response to increasing the negative IMF $Bz$ as demonstrated in Figure 10 and described in the Equation (4.5). The wide spreading of the data points in Figure 10 indicate to strong contribution of other effects controlling the shape of the dayside magnetopause such so the dependence of the "perigee" longitude on the IMF $Bz$ can be considered as very approximate and preliminary.

As we mentioned in the previous section the existence of two horizontal branches on the envelope boundary can be interpreted in terms of two asymptotic regimes of the magnetopause formation. The regime of large positive $Bz>20$ nT is attributed to the classical pressure balance [*Chapman and Ferraro,* 1931; *Spreiter et al.,* 1966] which is independent on the IMF $Bz$. Moreover it was assumed that one can neglect the contribution of the solar wind magnetic and thermal pressures to the pressure balance in the MP nose point because these pressures are usually less than one tenth of nPa. We have to emphasize that for the moderate northward IMF (positive $Bz<20$ nT) the MP nose point distance still depends significantly on the IMF $Bz$. We attribute this feature to a reconnection on the low latitude dayside magnetopause due to effect of the IMF $By$.

The dayside reconnection on low altitudes is controlled mainly by the IMF *Bz* and *By* components. This effect is neglected in most existing MP models. However DS00 model using the ANN technique for the MP data treatment introduces the IMF *By* component as important parameter controlling the dayside magnetopause dynamics. We have demonstrated the *By* effect on Plot 2. Considering Table 4 we can also find several examples of the GMCs accompanied by large absolute value of the *By* especially for small negative and positive *Bz* (>5 nT). Apparently the *By* effect is partially hidden by variations of the GMC location. However in 6 of 10 GMCs observed in 3 hour vicinity of the noon and under *Bz*>-5 nT the absolute value of the *By* is large than *Bz* indicating on substantial contribution of the *By* into reconnection on the dayside MP.

For the large southward IMF the boundary enveloping the necessary SW conditions for GMC is characterized by so called "*Bz* influence saturation", one of the newest MP effects which is modeled and intensively discussed during last decade [*Kuznetsov and Suvorova,* 1994; 1996b; *Shue et al.*, 1998; 2001; *Dmitriev and Suvorova*, 2000; *Dmitriev et al.*, 2001b; *Yang et al.,* 2003]. As we can see from Figure 1 and Table 1 different MP models indicate various SW dynamic pressures (from ~ 3 nPa to ~7 nPa) corresponding to *Bz* influence saturation for the subsolar MP point crossing the geosynchronous orbit. In the present study we obtain *Psw*=4.8 nPa for the GMC in the "perigee" point.

There is no complete physical explanation suggested for the *Bz* influence saturation as well as for the MP dawn-dusk asymmetry. One of the most probable phenomena responsible for the dawn-dusk asymmetry was suggested the asymmetrical storm-time ring current with maximum in the evening sector developing under strong southward IMF *Bz* [*McComas et al.*, 1993; *Itoh and Araki,* 1996]. Due to asymmetrical ring current [*Cummings,* 1966; *Burton et al.,* 1975] the dusk side of the magnetosphere where the ring current is maximal should be larger than dawn side. However during a geomagnetic storm another magnetospheric currents such as the tail current and the field align currents are intensified and significantly contribute to the magnetic field in the dayside magnetosphere [see *Alexeev et al.,* 2000; *Maltsev, and Lyatsky*, 1975]. Their relative contribution is different and unambiguous especially during strong geomagnetic storms.

Besides magnetic effect of the ring current we can consider a thermal pressure of the magnetospheric plasma $P_{tm}$ which is also contributed by ring current particles. Direct measurements of the thermal plasma in the magnetosphere [*Frank,* 1967; *Lui et al.,* 1987; *Lui and Hamilton,* 1992] show that the perpendicular pressure in the dayside region of geosynchronous orbit is about 1~2 nPa for quite geomagnetic conditions and it growths up to 4 nPa during strong geomagnetic storms. Such pressure is comparable with the solar wind pressure in the "regime of saturation" $Psw$=4.8 nPa. Therefore for large negative IMF $Bz$ the magnetosphere plasma pressure contribution to the magnetopause pressure balance can not be neglected. As result we can expand the pressure balance Equation (1.1) in the MP nose region:

$$k \cdot Pd = \frac{(2fB)^2}{8\pi} + P_{tm} \qquad (5.1)$$

Here we neglect the SW thermal and magnetic pressures. Compression and erosion affect the magnetopause magnetic field in different manner, so there are observed two different types of GMC events based on the morphology of the magnetic field signatures [*Rufenach et al.,* 1989; *Itoh and Araki,* 1996]. Using this fact *Kuznetsov and Suvorova* [1998b] further concluded that the erosion on the dayside magnetopause under strong negative $Bz$ leads to decrease of the coefficient $f$ in Equation (1.1) down to 0.5. In other words, the geomagnetic field pressure contribution in Equation (5.1) decreases and relative importance of the thermal pressure $P_{th}$ growths up. Under strong negative $Bz$ the magnetopause moves earthward due to reconnection which causes penetration of the IMF on smaller distances. However, the IMF influence can be terminated by a force of non-magnetic nature such as thermal pressure of the magnetospheric plasma. Hence we can suggest that the "$Bz$ influence saturation" is owing to magnetospheric thermal pressure contribution to the pressure balance on the dayside magnetopause.

The dawn-dusk asymmetry can be also interpreted in terms of the asymmetrical ring current plasma pressure. The magnetosphere plasma pressure excess in the evening pushes the magnetopause outward the Earth such that the magnetopause flank is "shifted" toward the dusk. According to the last investigation by *McComas et al.* [1993] of the local time distribution of the geosynchronous plasmasphere observation, at geomagnetically active times ($Kp>2$) the plasmospheric bulge moves from ~1600 LT toward noon with increasing activity. Thus, contribution in thermal pressure from both cold, dense plasmaspheric and high-energy ring current populations is larger in the afternoon portion of the magnetosphere than in the morning portion.

A representation of the MP shape with dawn-dusk asymmetry was suggested by *Kuznetsov and Suvorova* [1997, 1998b]. Figure 11 illustrates the magnetopause shifted on $dY$=2 Re toward the dusk (thick solid line) and crossing the geosynchronous orbit (black circle). The "non-shifted" magnetopause is indicated by dashed line. One can see that the "perigee" point (black circle) where the GMC is occurred is shifted toward the dawn while the MP nose point is shifted toward the dusk and located at larger distance. In other words, the SW conditions for the GMC in the MP nose point should be stronger than the necessary solar wind conditions for GMC.

The difference between the SW conditions required for GMC in the "perigee" and nose points depends on the magnetopause shift $dY$, which is associated with the ring current asymmetry, and on MP flaring which depends on the IMF $Bz$. Using the KS98 model we can roughly estimate that for $Bz$=-30 nT and magnetopause shift $dY$=2 Re the SW pressure required to push the MP nose point on distance 6.6 Re should be larger on 50% than the necessary SW pressure for GMC in the "perigee" point ($P_{sw}$=4.8 nPa). Hence the GMC in the MP nose point requires $P_{sw}$~7 nPa for large negative $Bz$. Such a value of the SW pressure is in good agreement with our suggestion on the magnetosphere plasma pressure contribution into the pressure balance in the MP nose point. Indeed estimation of the geodipole magnetic field energy (assuming $f$=0.5) in geosynchronous orbit gives us a value of geomagnetic field pressure of about 4.6 nPa (the first term in eq. 5.1). Our estimation of the SW dynamic pressure required for GMC in the nose point gives P$sw$=7 nPa. The difference of about 2 nPa can be attributed to the thermal pressure of the magnetospheric plasma (the second term in eq. 5.1) that is in a good agreement with experimental measurements [*Lui et al.,* 1987]. More precise determination of the SW conditions for the MP nose point will be possible after careful investigation of the MP dawn-dusk asymmetry dependence on the geomagnetic and solar wind conditions (IMF $By$ and $Bz$). Solution of this problem is a subject of our following study.

Finally we can apply the above results for estimation of validity of the MP models presented in Figure 1 and Table 1. We can certainly say that for large positive IMF $Bz$ when the dawn-dusk asymmetry is very weak the most accurate prediction of the SW conditions for GMC is provided by the KS98 model. Indeed this model predicts for GMC the solar wind dynamic pressure $Pd$=24.5 nPa which is mostly close to the necessary SW pressure $Psw$=21 nPa obtained here for the large positive $Bz$. Estimation of the most plausible MP model for moderate positive and negative $Bz$ is difficult. We still do not know definitely the actual dynamics of the magnetopause size and shape during strong geomagnetic and interplanetary disturbances accompanied by intensive erosion on the dayside magnetopause. In the present study we have found only the minimal level of the SW disturbance characterized by the SW pressure and IMF $Bz$. Our prediction is valid for the "perigee" point which is not always coincident with the MP nose point. Moreover, the difference between locations of the "perigee" and nose points varies with SW conditions and dipole tilt angle and, thus, can not be simply accounted. Consideration of Figure 1 shows that for the large negative $Bz$ the Ch02 and Sh98 models predict the SW dynamic pressure required for GMC in the MP nose point ($Pd$=6.3 and $Pd$=7.4 respectively) very close to our estimation ($P_{sw}$~7). However the Ch02 and Sh98 models do not represent the dawn-dusk asymmetry and, thus, the MP nose point is the nearest to the Earth in these models. Hence the Ch02 and Sh98 models do not predict the GMC when the SW dynamic pressure less than $Pd$=6.3 and $Pd$=7.4 respectively. This is one of possible reasons of the 0.5 Re uncertainty incorporated in the Sh98 model to predict the GMCs [*Shue et al.*, 2000b, 2001]. Hence we can only indicate here that the PR96, KS98 and DS00 models are more plausible because they are able to predict the GMCs under relatively weak SW pressures $Psw$~5 required for GMC. But the PR96 model does not predict the effect of "$Bz$ influence saturation" and therefore this model significantly underestimates the MP distances under IMF $Bz$ less that –10 nT. Strictly saying there is no MP model accurately predicting the GMCs for southward IMF. Modeling the dayside magnetopause under strongly disturbed solar wind conditions requires consideration of different equally important effects such as "*Bz* influence saturation", dawn-dusk asymmetry and tilt angle variation. Development of realistic MP model for prediction of the GMCs is a next step in our work.

## 6. Summary and Conclusions

1. Analysis of the locations of geosynchronous satellites GOESs and LANLs in the realistic geomagnetic field shows that the longitudinal and latitudinal effects are substantial (up to 20%). These effects have strong influence on the prediction capability of modern empirical models of the magnetopause.

2. Effective method for accurate matching of the upstream solar wind conditions with observations of near earth satellites is suggested. The method is based on 1 min *Dst* (SYM) index as a relevant indicator of the solar wind pressure influence on the magnetosphere.

3. The alternative method for GMC identification is suggested for the MPA LANL key parameters when the plasma spectra are not available.

4. The data set of 638 magnetopause crossings and 5866 magnetosheath measurements by 8 geosynchronous satellites in 1995 to 2001 is carefully prepared.

5. From analysis of the GMC data set the necessary conditions in the solar wind total pressure *Psw* and IMF *Bz* required for GMC are determined and described numerically by hyperbolic tangent function (Equation 4.3) containing two asymptotes at $Psw=21$ nPa ($Bz>>0$) and $Psw=4.8$ nPa ($Bz<<0$) where the MP location does not depend on the IMF *Bz* value.

6. Under necessary solar wind conditions the magnetopause crosses the geosynchronous orbit in the so-called "perigee" point which location varies significantly in aGSM longitudes and latitudes. Longitudinal variation of the "perigee" point location is associated with the MP dawn-dusk asymmetry. It is found that aGSM longitude of the "perigee" point tends to shift toward the morning sector with strong southward IMF Bz (Equation 4.4). Latitudinal variations of the "perigee" point are revealed being controlled by the dipole tilt angle. This dependence is approximated by linear function with a slope of about –0.5 (Equation 4.5).

7. We propose that the both effects, *Bz* influence saturation and dawn-dusk asymmetry, can be explained by properties of the thermal plasma distribution in the disturbed magnetosphere. The saturation of *Bz* influence under strong negative *Bz* can be attributed to relatively strong contribution of the magnetosphere thermal pressure into the pressure balance on the magnetopause when the reconnection/erosion processes are active. The dawn-dusk asymmetry can be explained by enhancement of plasma pressure due to asymmetrical storm-time ring current. The ring current excess at evening hours pushes the magnetopause such that the distance to the MP in the dusk is higher than the distance in the dawn.

8. System of the Equations (4.3), (4.4) and (4.5) represents an empirical model which permits predicting location of the "perigee" point and solar wind conditions required for magnetopause crossing of the geosynchronous orbit in this point.

**Acknowledgement**

This work is supported by grant NSC-91-2811-M-008-019. We are very grateful to Prof. S.N. Kuznetsov for valuable recommendation in methodic of the magnetopause crossing selection.

# References

Alexeev, I.I. E.S. Belenkaya, and C.R. Clauer, A model of region 1 field-aligned currents dependent on ionospheric condactivity and solar wind parameters, *J. Geophys. Res.*, *103*, 21,119, 2000.

Boardsen S.A., T.E. Eastman, T. Sotirelis, and J.L. Green, An empirical model of the high-latitude magnetopause, *J. Geophys. Res., 105*, 23193-23219, 2000.

Burton, R.K., R.L. McPherron, and C.T. Russell, An empirical relationship between interplanetary conditions and *Dst*, *J. Geophys. Res., 80*, 4204, 1975.

Cummings, W.D., Asymmetric ring current and the low-latitude disturbance daily variation. *J. Geophys. Res., 71*, 4495, 1966.

Chao, J.K., D.J. Wu, C.-H. Lin, Y.-H. Yang, X.Y.Wang, M.Kessel, S.H.Chen, and R.P.Lepping., Models for the size and shape of the Earth's magnetopause and bow shock, in *Space Weather Study Using Multipoint Techniques , Cospar Colloq. Ser.*, vol.12, edited by L.-H. Lyu, pp.127-134, Pergamon, Elsevier Science Ltd., 2002.

Chapman, S., and V.C.A.Ferraro, A new theory of magnetic storms, 1, The initial phase (continued), *J. Geophys. Res., 36*, 171-186, 1931.

Collier, M.R., J.A. Slavin, R.P. Lepping, A. Szabo, and K. Ogilvie, Timing accuracy for the simple planar propagation of magnetic field structures in the solar wind, *Geophys. Res. Lett., 25*, 2509, 1998.

Degtyarev, V.I., O.I. Platonov, G.V.Popov, Longitudinal effect in particle spectra changes measured at geostationary satellites (in Russian), *Geomagn. Aeron., 25* (6), 1002-1004, 1985.

Dmitriev, A.V., and Yu.V. Orlov, NeuroShell and multifactor analysis coupling in the solution of some space physics problems, paper presented at International Workshop and Symposium on Statistical Physics, Academia Sinica, Taipei, Taiwan, R.O.C., Aug. 3-11, 1997.

Dmitriev, A.V., and A.V. Suvorova, Three-dimensional artificial neural network model of the dayside magnetopause, *J. Geophys. Res., 105*, 18909, 2000a.

Dmitriev A.V., and A.V. Suvorova, Artificial neural network model of the dayside magnetopause: Physical consequences, *Phys. Chem. Earth*, 25(1/2), 169-172, 2000b.

Dmitriev A. V. and A. V. Suvorova (2004), Geosynchronous Magnetopause Crossings on October 29-31, 2003, *Cosmic Research*, 42(6), 551-560.

Dmitriev, A.V., Yu.V. Orlov, I.V. Persiantsev, and A.V. Suvorova, 3D dayside magnetopause model with artificial neural networks, *Geomagn. Aeron.*, Engl. Transl., *39* (5), 86-95, 1999.

Dmitriev A.V., J.K. Chao, and Y.-H. Yang, Saturation of the *Bz*-influence for the geosynchronous magnetopause crossings, paper presented at Taiwan Geophysical Meeting, Taipei, September 25-27, 2001a.

Dmitriev A.V., J.K. Chao, Y.-H. Yang, and A.V. Suvorova, Magnetopause dawn–dusk asymmetry at the geosynchronous orbit, in Proc. of Taiwan Geophysical Meeting, Taipei, September 25-27, 2001b.

Dmitriev, A.V., J.K. Chao, Y.H. Yang, C.-H. Lin, and D.J. Wu, Possible Sources of the Difference between a Model Prediction and Observations of Bow Shock Crossings, *TAO*, *13*, 4, 499, 2002a.

Dmitriev A., J. Chao, Y. Yang, A. Suvorova, Dawn-Dusk Asymmetry of the Earth's Magnetosphere, in Proc. of International Conference of Earth System Science, Chungli, Taiwan, June 3-4, S3-S7, 2002b.

Dmitriev A., J.-K. Chao, D.-J. Wu, Comparative study of bow shock models using Wind and Geotail observations, *J. Geophys. Res.*, 108(A12), 1464, doi:10.1029/2003JA010027, 2003.

Dmitriev A.V., A.V. Suvorova, J.-K. Chao, Y.-H. Yang (2004), Dawn-dusk asymmetry of geosynchronous magnetopause crossings, *J. Geophys. Res.*, 109, A05203 doi: 10.1029/2003JA010171, 2004.

Dmitriev A., J.-K. Chao, M. Thomsen, A. Suvorova (2005a), Geosynchronous magnetopause crossings on October 29-31, 2003, *J. Geophys. Res.*, J. Geophys. Res., 110(A8), A08209, doi:10.1029/2004JA010582.

Dmitriev A., J.-K. Chao, A. Suvorova, K. Ackerson, K. Ishisaka, Y. Kasaba, H. Kojima, H. Matsumoto (2005b), Indirect estimation of the solar wind conditions in 29-31 October 2003, *J. Geophys. Res.*, 110, A09S02, doi:10.1029/2004JA010806.

Dmitriev A., N. Crosby, J.-K. Chao (2005c), Interplanetary sources of space weather disturbances in 1997 to 2000, *Space Weather*, *3*(3), S03001, 10.1029/2004SW000104.

Dmitriev, A. V., P. T. Jayachandran, and L.-C. Tsai (2010), Elliptical model of cutoff boundaries for the solar energetic particles measured by POES satellites in December 2006, *J. Geophys. Res.*, 115, A12244, doi:10.1029/2010JA015380.

Dmitriev, A. V., Suvorova, A. V., J.-K. Chao (2011), A predictive model of geosynchronous magnetopause crossings, *J. Geophys. Res.*, 116, A05208, doi:10.1029/2010JA016208.

Frank, L.A., On the extraterrestrial ring current during geomagnetic storms, *J. Geophys. Res.*, *72*, 3753-3767, 1967.

Grafodatskii, O.S. et al., On longitudinal variations of the energetic electron fluxes in vicinity of the geostationary satellites (in Russian), *Issled. geomag., aeron. i physike of the sun*, *85*, pp.3-7, Nauka, Moscow, 1989.

Holzer R.E, and J.A. Slavin, Magnetic flux transfer associated with expansions and contractions of the dayside magnetosphere, *J.Geophys.Res., 83,* 3831, 1978.

Itoh K., and T. Araki, Analysis of geosynchronous magnetopause crossings, in *Proceedings of Solar Terrestrial Predictions Workshop*, Hitachi, Japan, Jan.23-27, pp.26-29, 1996.

Karimabadi, H., T. B. Sipes, H. White, M. Marinucci, A. Dmitriev, J. K. Chao, J. Driscoll, and N. Balac (2007), Data mining in space physics: MineTool algorithm, *J. Geophys. Res.*, 112, A11215, doi:10.1029/2006JA012136.

Kuznetsov S.N., and A.V. Suvorova, Solar wind control of the magnetopause shape and location, paper presented at International workshop "Space Radiation Environment: empirical and physical models", Dubna, June 5-7,1993.

Kuznetsov S.N., and A.V. Suvorova, Influence of solar wind to some magnetospheric characteristics, in *Proceedings of WDS'94*, *PartII-Physics of plasmas and ionized media*, edited by J.Safrankova, pp.116-123, Charles University, Prague, 1994.

Kuznetsov, S.N., and A.V. Suvorova, Solar wind control of the magnetopause shape and location, *Radiat. Meas., 26*, 413, 1996a.

Kuznetsov S.N., and A.V. Suvorova, On two regimes of solar wind interaction with magnetosphere, paper presented at First EGS Alfven Conference on Low-Altitude Investigation of Dayside Magnetospheric Boundary Processes, Kiruna, Sweden, Sep.9-13, 1996b.

Kuznetsov, S.N., and A.V. Suvorova, Magnetopause shape near geosynchronous orbit (in Russian), *Geomagn. Aeron.*, *37*, 1, 1997.

Kuznetsov, S.N., and A.V. Suvorova, An empirical model of the magnetopause for broad ranges of solar wind pressure and Bz IMF, in *Polar cap boundary phenomena*, *NATO ASI Ser.*, edited by J.Moen, A.Egeland and M.Lockwood, pp.51-61, Kluwer Acad., Norwell, Mass., 1998a.

Kuznetsov, S.N., and A.V. Suvorova, Solar wind magnetic field and plasma during magnetopause crossings at geosynchronous orbit, *Adv. Space Res., 22*(1), 63-66, 1998b.

Kuznetsov, S.N., A.V. Suvorova, and A.V. Dmitriev, Magnetopause shape and size: Relation with parameters of the interplanetary medium, *Geomagn. Aeron.*, Engl. Transl., *38* (6), 7-16, 1998.

Kuznetsov, S.N., A.V. Suvorova, G.N. Zastenker, and D.G. Sibeck, Solar wind control of the geomagnetopause position, in *Proceedings of 1992 STEP Symposium, Cospar Colloq. Ser.,* vol.5, pp. 293-296, Elsevier Sci. Inc., New York, 1994.

Kuznetsov, S.N., G.N. Zastenker, and A.V. Suvorova, Correlation between interplanetary conditions and the dayside magnetopause, *Cosm. Res., 30*, 466, 1992.

Lui, A.T.Y., and D.C. Hamilton, Radial profiles of quiet time magnetospheric parameters, *J. Geophys. Res., 97*, 19,325-19,332, 1992.

Lui, A.T.Y., R.W. McEntrie, and S.M. Krimings, Evolution of the ring current during two geomagnetic storms, *J. Geophys. Res., 92*, 7459-7470, 1987.

Maltsev, Y.P., and W.B. Lyatsky, Field-aligned currents and erosion of the dayside magnetopause, *Planet. Space. Sci., 23*, 1237-1260, 1975.

McComas, D.J., et al., Magnetospheric plasma analyzer: Initial three-spacecraft observations from geosynchronous orbit, *J. Geophys. Res.*, *98*, 13453-13465, 1993.

McComas, D.J., R.C.Elphic, M.B.Moldwin, and M.F.Thomsen, Plasma observations of magnetopause crossings at geosynchronous orbit, *J. Geophys. Res.*, *99*, 21249-21255, 1994.

Merka, J., A. Szabo, T.W. Narock, J.H. King, K.I. Paularena, and J.D. Richardson, A comparison of IMP 8 observed bow shock positions with model predictions, *J. Geophys. Res., 108*(A2), 1077, doi:10.1029/2002JA009384, 2003.

Petrinec, S.M., and C.T. Russell, An examination of the effect of dipole tilt angle and cusp regions on the dayside magnetopause, *J. Geophys. Res., 100*, 9559-9566, 1995.

Petrinec, S.M., and C.T. Russell, Near-Earth magnetotail shape and size as determined from the magnetopause flaring angle, *J. Geophys. Res., 101*, 137, 1996.

Přech L., J. Šafránková, Z. Němeček, K. Kudela, M. Slivka (2005), Plasma flow variations and energetic protons upstream of the earth's bow shock: A statistical study, Advances in Space Research, 36(12), 2345-2350.

Richardson, J.D., and K. Paularena, Plasma and Magnetic Field Correlations in the Solar Wind, *J. Geophys. Res., 106*, 239, 2001.

Riazantseva, M.O., P.A. Dalin, A.V. Dmitriev, Yu.V. Orlov, K.I. Paularena, J.D. Richardson and G.N. Zastenker, A multifactor analysis of parameters controlling solar wind ion flux correlations using an artificial neural network technique, *J. Atmos. Sol. Terr. Phys., 64*, 657, 2002.

Roelof, E.C., and D.G. Sibeck, The magnetopause shape as a bivariate function of IMF $B_z$ and solar wind dynamic pressure, *J. Geophys. Res., 98*, 21,421-21,450, 1993.

Rufenach C.L., Jr.,R.F., Martin, and H.H. Sauer, A study of geosynchronous magnetopause crossings, *J. Geophys. Res., 94*, 15125, 1989.

Russell, C.T., M. Ginskey, and S.M. Petrinec, Sudden impulses at low-latitude stations: Steady state response for northward interplanetary magnetic field, *J. Geophys. Res., 99*, 253, 1994a.

Russell, C.T., M. Ginskey, and S.M. Petrinec, Sudden impulses at low-latitude stations: Steady state response for southward interplanetary magnetic field, *J. Geophys. Res., 99*, 13403, 1994b.

Schield, M.A., Pressure Balance between solar wind and magnetosphere, *J. Geophys. Res., 74*, 1275-1286, 1969.

Shue, J.-H., J.K. Chao, H.C. Fu, C.T. Russell, P. Song, et al., A new functional form to study the solar wind control of the magnetopause size and shape, *J. Geophys. Res.*, *102*, 9497, 1997.

Shue, J.-H., P. Song, C.T. Russell, J.T. Steinberg, J.K. Chao et al., Magnetopause location under extreme solar wind conditions, *J. Geophys. Res.*, *103*, 17691, 1998.

Shue, J.-H, C.T. Russell, and P. Song, Shape of the low-latitude magnetopause: Comparison of models, *Adv. Space Res.*, *25*(7/8), 1471, 2000a.

Shue, J.-H, P. Song, C.T. Russell, J.K. Chao, and Y.-H.Yang, Toward predicting the position of the magnetopause within the geosynchronous orbit, *J. Geophys. Res.*, *105*, 2641-2656, 2000b.

Shue, J.-H., P. Song, C.T. Russell, M.F. Thomsen, and S.M. Petrinec, Dependence of magnetopause erosion on southword interplanetary magnetic field, *J. Geophys. Res.*, *106*, 18,777-18,788, 2001.

Song, P., et al., Polar observations and model predictions during May 4, 1998, magnetopause, magnetosheath, and bow shock crossings, *J. Geophys. Res.*, *106*, 18,927-18,942, 2001.

Sotirelis, T., and C.-I. Meng, Magnetopause from pressure balance, *J. Geophys. Res., 104*, 6889, 1999.

Spreiter, J.R., A.L. Summers, and A.Y. Alksne, Hydromagnetic flow around the magnetosphere, *Planet. Space Sci., 14*, 223-253, 1966.

Suvorova, A.V., A.V. Dmitriev, and S.N. Kuznetsov, Dayside magnetopause models, *Radiat. Meas., 30*, 687, 1999.

Suvorova A., A. Dmitriev, J.-K. Chao, M. Thomsen (2005), Y.-H. Yang, Necessary conditions for the geosynchronous magnetopause crossings, *J. Geophys. Res.*, 110, A01206, doi:10.1029/2003JA010079.

Tsyganenko, N. A., A model of the near magnetosphere with a dawn-dusk asymmetry 1. Mathematical structure, *J. Geophys. Res.*, *107* (A8), doi: 10.1029/2001JA000219, 2002.

Tsyganenko, N. A., A model of the near magnetosphere with a dawn-dusk asymmetry 2. Parameterization and fitting to observations, *J. Geophys. Res., 107*(A8), doi: 10.1029/2001JA000220, 2002.

Weimer, D.R., D.M. Ober, N.C. Maynard, W.J. Burke, M.R. Collier, Variable time delays in the propagation of the interplanetary magnetic field, *J. Geophys. Res.*, *107*(A8), 10.1029/2001JA009102, 2002.

Wrenn, G.L., J.F.E. Johnson, A.J. Norris, and M.F. Smith, GEOS-2 magnetopause encounters: Low energy (<500 eV) particle measurements, *Adv. Space Res.*, *1,* 129, 1981.

Yang, Y.-H., et al., Comparison of three magnetopause prediction models under extreme solar wind conditions, *J. Geophys. Res.*, *107*(A1), 10.1029/2001JA000079, 2002.

Yang, Y.-H., J. K. Chao, A. V. Dmitriev, C.-H. Lin, and D. M. Ober, Saturation of IMF Bz influence on the position of dayside magnetopause, *J. Geophys. Res.*, *108*(A3), 1104, doi:10.1029/2002JA009621, 2003.

Veselovsky, I.S., A.V. Dmitriev, and A.V. Suvorova, Average parameters of the solar wind and interplanetary magnetic field at the Earth's orbit for the last three solar cycles, *Sol. Sys. Res.*, Engl. Transl., *32*(4), 310, 1998.

**Figure captions**

Fig. 1. Solar wind dynamic pressure and IMF *Bz* predicted for magnetopause subsolar distance 6.6 Re by the PR96, Sh98, KS98, DS00 and Ch02 models. The model predictions are indicated, respectively, by black solid, dashed, dashed dotted, dotted (gray for IMF *By*=0 nT and black for IMF *By*=20 nT) and gray solid lines.

Fig. 2. Magnetic field of internal sources calculated in the geographic equator on distance 6.6 Re from dipole approach (dashed curve) and IGRF model (solid curve). Vertical lines indicate longitude and corresponding magnetic field for different geosynchronous satellites.

Fig. 3. Distribution of occurrence probabilities in GSM coordinates latitude versus local time for different geosynchronous orbits: GOES-8 (a), GOES-9 and GOES-10 (b), LANL-1990 (c), LANL-1991 (d), LANL-1994 (e), LANL-1997 (f) and LANL-1989 (g). The occurrence probability is represented in gray scale and varies from ~$10^{-4}$ (light gray) to ~$10^{-2}$ (black).

Fig. 4. Distribution of total geomagnetic field calculated from T01 model in GSM coordinates for the dayside sector under different conditions: (a) *Pd*=4 nPa, *Bz*=0 nT, *Dst*=0 nT, *PS*=0°; (b) *Pd*=9 nPa, *Bz*=0 nT, *Dst*=0 nT, *PS*=0°; (c) *Pd*=2 nPa, *Bz*=-10 nT, *Dst*=-100 nT, *PS*=0° (d) *Pd*=4 nPa, *Bz*=0 nT, *Dst*=0 nT, *PS*=20°. The other model parameters are accepted as *By*=5 nT, G1=6 and G2=10.

Fig. 5. Scatter plot of the GMCs in aGSM coordinates. Magnetopause crossings by GOES and LANL, and magnetosheath measurements are indicated by black crosses, gray crosses and triangles respectively.

Fig. 6. Scatter plot Scatter plot of the GMCs in space of the solar wind parameters SW pressure versus aGSM *Bz*. Magnetopause crossings by GOES and LANL, and magnetosheath measurements are indicated by black crosses, gray crosses and triangles respectively. The thick curve is expressed by Eq. (4.3).

**Fig. 7.** Two-dimensional distribution of occurrence number of the GMC intervals binned in the coordinates $P_{sw}$ (in logarithmic scale) versus IMF $B_z$ in aGSM.

**Fig. 8.** Approximation of the meaningful GMCs and magnetosheath measurements from the down edge bins (Figure 6) by hyperbolic tangent in the space of parameters SW pressure and IMF $B_z$.

**Fig. 9.** Scatter plot of the magnetosheath measurements (triangles) and entrances by GOES (crosses) and LANL (asterisks) from 20% vicinity of the envelope boundary in aGSM coordinates latitude versus local time.

**Fig. 10.** Scatter plot of the magnetosheath entrances (crosses) from 20% vicinity of the envelope boundary in aGSM coordinates local time versus IMF $B_z$.

**Fig. 11.** Dependence of latitude of the GMCs (crosses) and magnetosheath measurements from 20% vicinity of the enveloping boundary on dipole tilt angle $PS$.

**Fig. 12.** KS98 model calculation for the magnetopause shifted on $dY=2$ Re toward the dusk (thick solid line) and "non-shifted" magnetopause (dashed line). The nose, "zero" and "perigee" points are indicated by circle, asterisk and triangle respectively.

**Plot. 1.** GMC identification using GOES-8 magnetic measurements and solar wind data from the ACE upstream monitor on April 18, 2000. See details in the text.

**Plot. 2.** GMC identification using LANL-1994 plasma measurements and solar wind data from the Wind upstream monitor on May 24, 2000. See details in the text.

**Plot. 3.** Observations of high-amplitude surface waves on the magnetopause by GOES-8, 10 and LANL-1991 on October 21-22, 2001. See details in the text.

**Plot. 4.** LANL-1991 observations of high and low energy plasma (density and components of velocity) during high-amplitude surface waves on October 21-22, 2001. See details in the text.

**Table 1. Solar wind dynamic pressure *Pd* (nPa) required for GMC.**

| Model | $Bz$=30 nT | $Bz$=-30 nT |
|---|---|---|
| PR96 | 30.3 | 0.7 |
| KS98 | 24.5 | 4 |
| SH98 | 39.3 | 7.4 |
| DS00 | 27~35 | 3~12 |
| CH02 | 45 | 6.3 |

**Table 2. Average Geographic Longitudes and Geomagnetic Field magnitudes corresponding to location of geosynchronous satellites.**

| Satellite name | Longitude | $H$ (nT) |
|---|---|---|
| GOES 8 (G8) | -75 | 108 |
| GOES 9 (G9) | -135 | 106 |
| GOES 10 (G10) | -135 | 106 |
| LANL-1990 (L0) | -45 | 105 |
| LANL-1991 (L1) | 0 | 102 |
| LANL-1994 (L4) | 135 | 113 |
| LANL-1997 (L7) | 75 | 110 |
| LANL-1989 (L9) | -165 | 107 |

**Table 3. Common statistics of the GMC intervals**

| Satellite | Case events | GMCs | Measurements in MS |
|---|---|---|---|
| GOES | 69 | 150 (148) | 3130 |
| LANL | 100 | 170 (170) | 2736 |
| TOTAL | 169 | 320 (318) | 5866 |

**Table 4. GMCs and corresponding conditions from vicinity of the envelope boundary**

| Date | UT | GS | UM | $LT$ (h) | $Lat$ (°) | $Psw$ (nPa) | $Bz$ (nT) | $By$ (nT) | $Dst$ (nT) |
|---|---|---|---|---|---|---|---|---|---|
| 10.3.1998 | 1754 | G9 | ACE | 9.703 | 3.64 | 5.63 | -15.90 | 4.82 | -110 |
| 10.3.1998 | 1759 | G9 | ACE | 9.648 | 3.32 | 5.48 | -15.10 | 1.39 | -114 |
| 4.5.1998 | 349 | L7 | Wind | 8.634 | -11.8 | 5.58 | -30.10 | -22.40 | -164 |
| 4.5.1998 | 520 | L7 | Wind | 9.901 | -18.5 | 15.30 | -0.51 | -34.50 | -264 |
| 16.7.1998 | 424 | L4 | Wind | 11.19 | -21.4 | 7.21 | -12.00 | -14.50 | -55 |
| 27.8.1998 | 306 | L4 | ACE | 10.06 | -7.2 | 6.35 | -8.80 | 7.29 | -154 |
| 27.8.1998 | 309 | L7 | ACE | 7.903 | -5.97 | 7.69 | -8.95 | 4.58 | -157 |
| 27.8.1998 | 316 | L4 | ACE | 10.22 | -7.26 | 6.90 | -9.83 | 3.30 | -155 |
| 27.8.1998 | 839 | L1 | Wind | 9.377 | -1.12 | 6.99 | -12.90 | 4.47 | -154 |
| 18.2.1999 | 1032 | L1 | ACE | 11.37 | 3.39 | 17.50 | 8.90 | 16.00 | -86 |
| 29.3.1999 | 626 | L7 | Geotail | 11.19 | -5.33 | 9.11 | -6.99 | -4.47 | 5 |
| 16.4.1999 | 1627 | G8 | ACE | 11.59 | -9.09 | 15.40 | 1.90 | 0.61 | 41 |
| 22.10.1999 | 505 | L7 | Wind | 10.11 | 3.99 | 4.80 | -31.10 | -7.44 | -171 |
| 8.6.2000 | 1320 | L1 | Wind | 13.81 | -22 | 19.90 | 17.90 | 7.45 | -8 |
| 13.7.2000 | 1136 | L1 | ACE | 11.47 | -22.2 | 19.70 | 9.34 | 9.63 | 31 |
| 12.8.2000 | 704 | L7 | ACE | 11.87 | -15.1 | 5.10 | -27.50 | -8.29 | -171 |
| 31.3.2001 | 255 | L4 | ACE | 9.601 | -5.42 | 20.70 | 17.30 | 30.00 | 50 |
| 31.3.2001 | 1655 | G8 | ACE | 12.39 | -1.23 | 4.83 | -29.20 | -12.90 | -240 |
| 8.4.2001 | 2056 | G0 | ACE | 12.48 | -6.01 | 14.60 | -0.66 | -2.06 | -47 |
| 8.4.2001 | 2057 | L1 | ACE | 10.55 | -13.2 | 14.20 | -1.80 | -2.47 | -49 |
| 8.4.2001 | 2101 | G0 | ACE | 12.55 | -5.73 | 14.10 | 1.46 | -1.54 | -48 |
| 25.9.2001 | 2137 | G0 | Geotail | 11.66 | 11.3 | 22.60 | 16.40 | -9.47 | -3 |
| 3.10.2001 | 1129 | L0 | ACE | 9.704 | 14.7 | 5.52 | -14.60 | -14.20 | -148 |
| 21.10.2001 | 1846 | G8 | Wind | 14.11 | 6 | 13.20 | -2.52 | 20.20 | -107 |
| 22.10.2001 | 1713 | G8 | Wind | 12.52 | 10.9 | 8.68 | -5.85 | 3.67 | -129 |

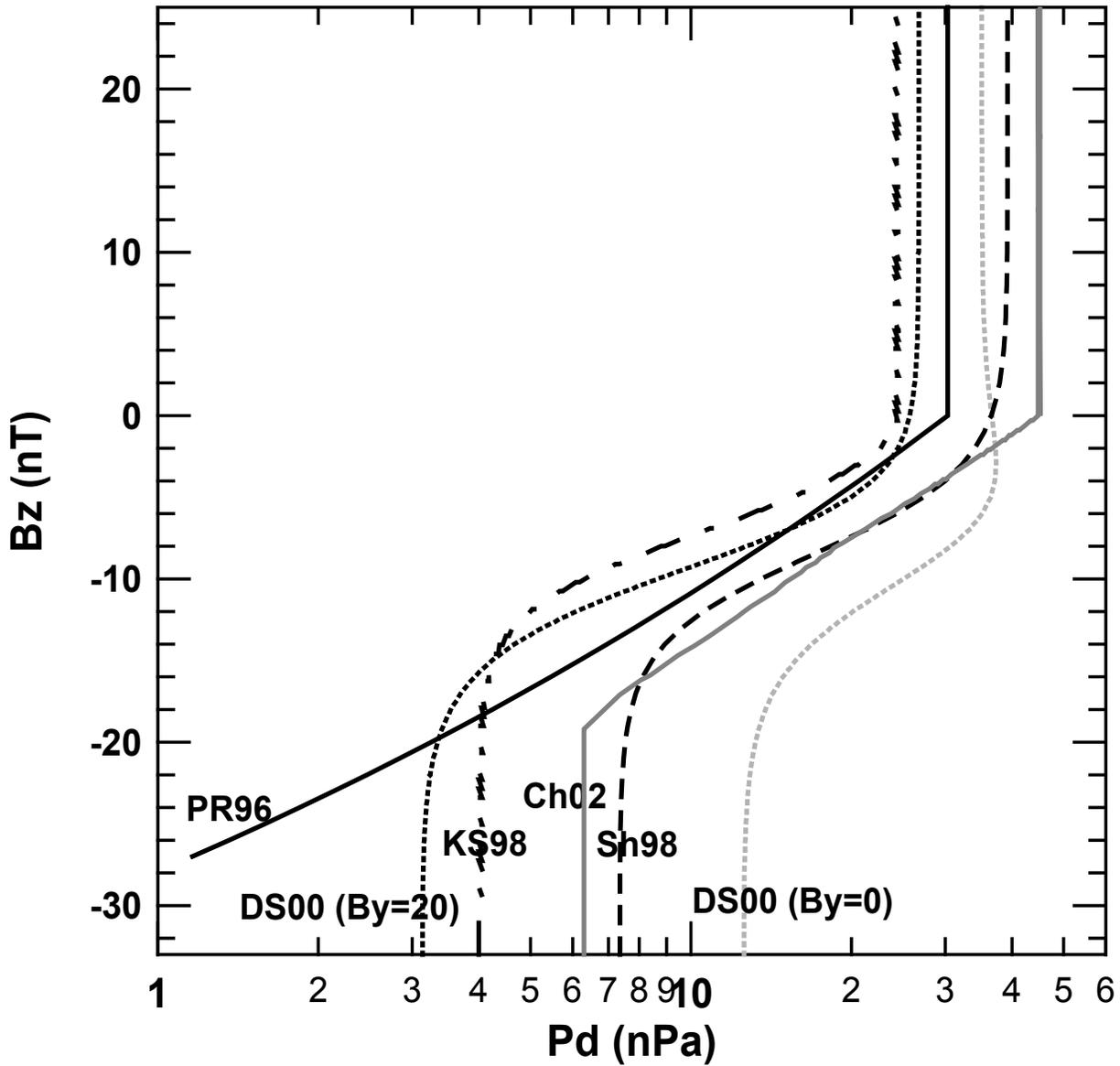

**Fig. 1.** Solar wind dynamic pressure and IMF *Bz* predicted for magnetopause subsolar distance 6.6 Re by the PR96, Sh98, KS98, DS00 and Ch02 models. The model predictions are indicated, respectively, by black solid, dashed, dashed dotted, dotted (gray for IMF *By*=0 nT and black for IMF *By*=20 nT) and gray solid lines.

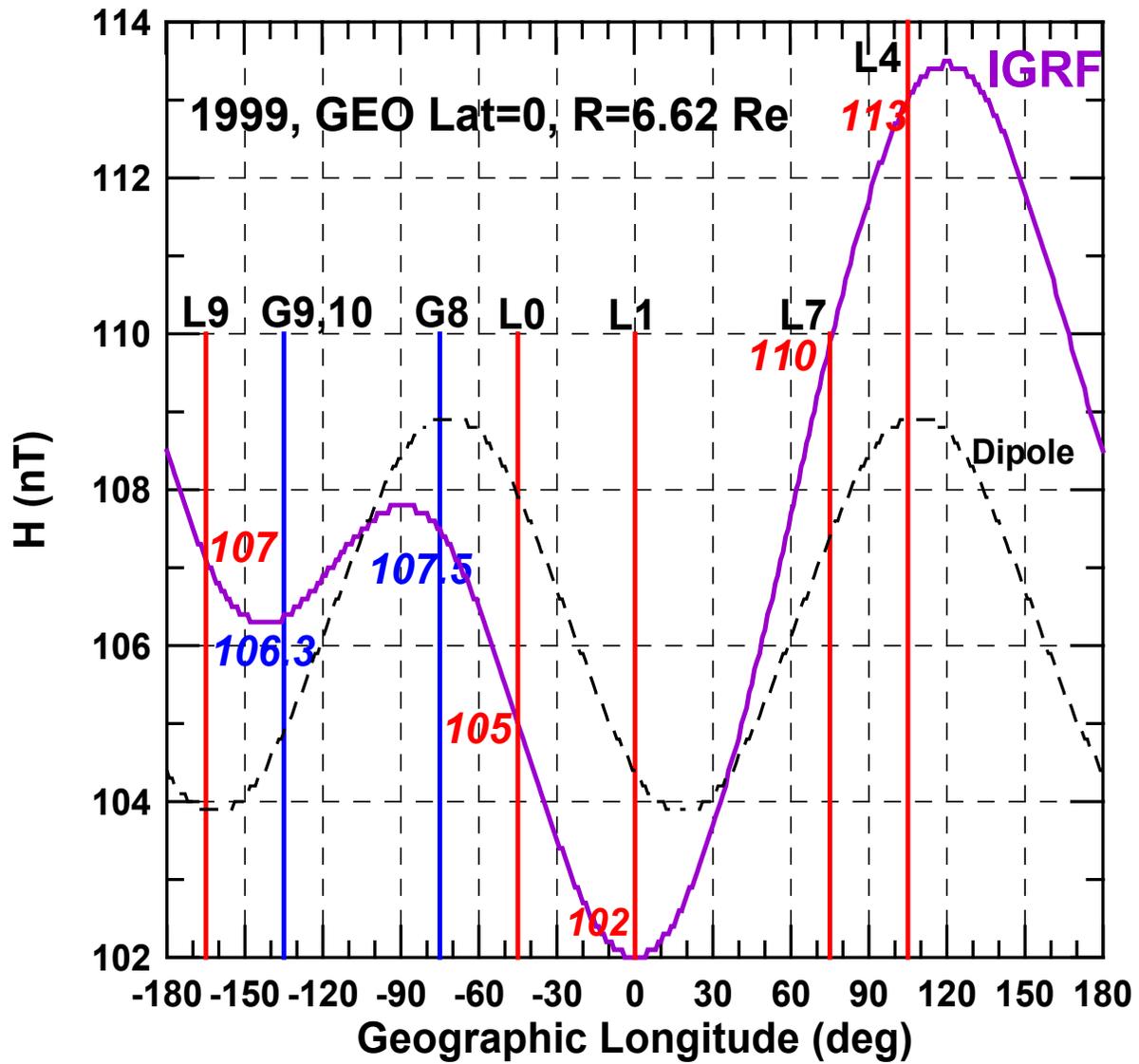

**Fig. 2.** Magnetic field of internal sources calculated in the geographic equator on distance 6.6 Re from dipole approach (dashed curve) and IGRF model (solid curve). Vertical lines indicate longitude and corresponding magnetic field for different geosynchronous satellites.

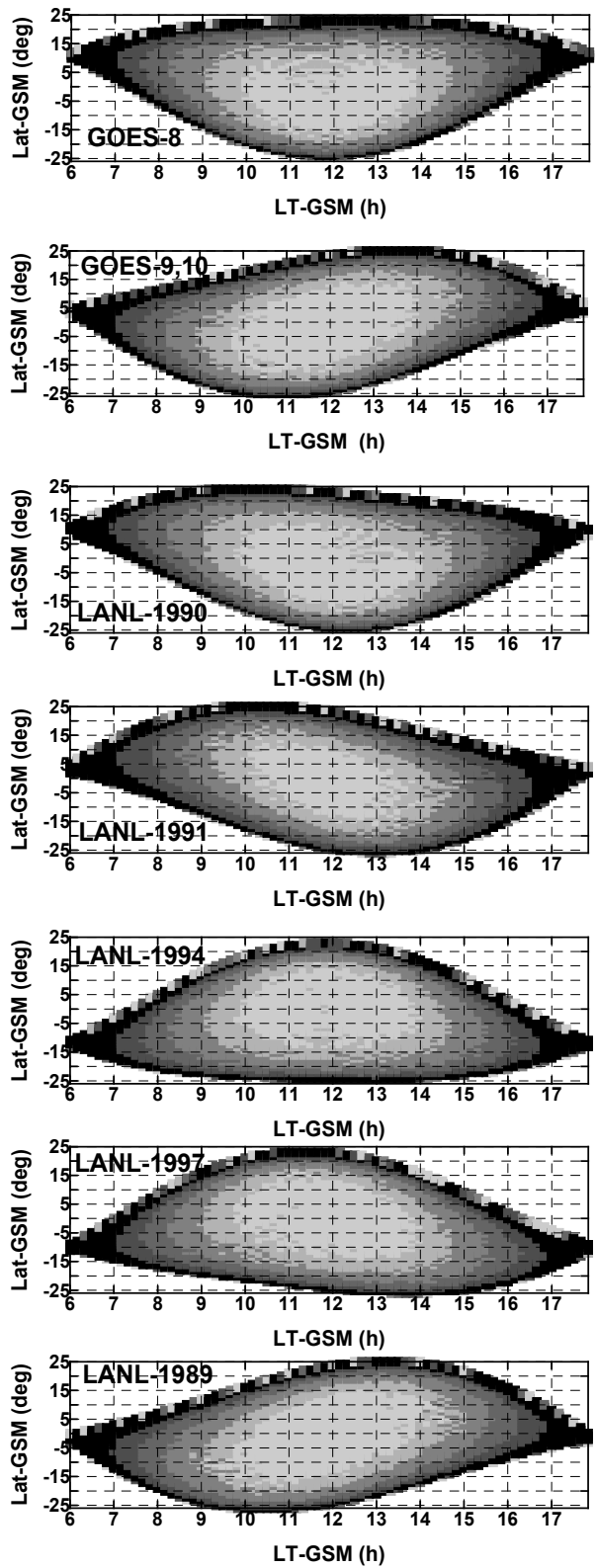

**Fig. 3.** Distribution of occurrence probabilities in GSM coordinates latitude versus local time for different geosynchronous orbits: GOES-8 (a), GOES-9 and GOES-10 (b), LANL-1990 (c), LANL-1991 (d), LANL-1994 (e), LANL-1997 (f) and LANL-1989 (g). The occurrence probability is represented in gray scale and varies from ~$10^{-4}$ (light gray) to ~$10^{-2}$ (black).

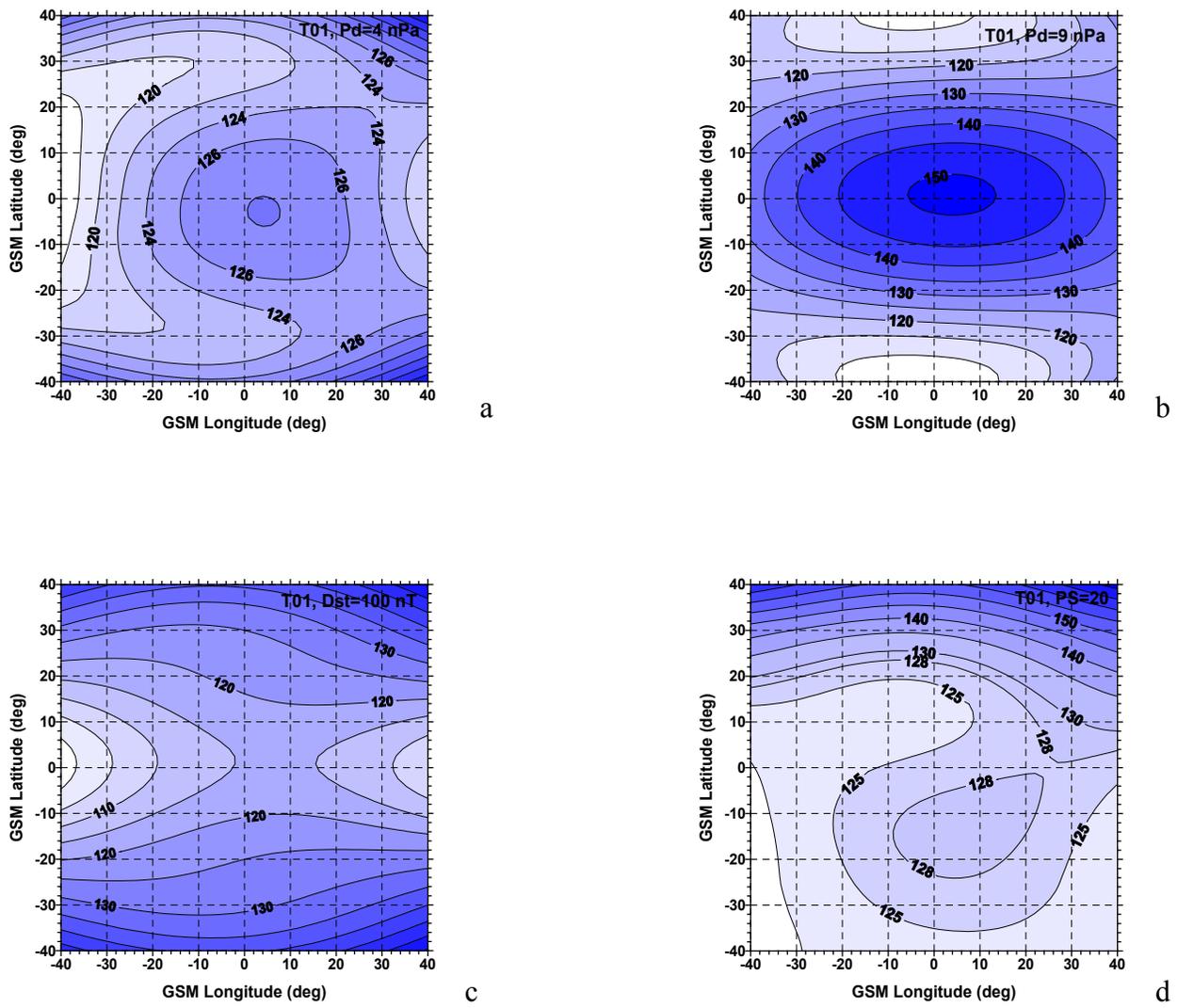

**Fig. 4.** Distribution of total geomagnetic field calculated from T01 model in GSM coordinates for the dayside sector under different conditions: (a) *Pd*=4 nPa, *Bz*=0 nT, *Dst*=0 nT, *PS*=0°; (b) *Pd*=9 nPa, *Bz*=0 nT, *Dst*=0 nT, *PS*=0°; (c) *Pd*=2 nPa, *Bz*=-10 nT, *Dst*=-100 nT, *PS*=0° (d) *Pd*=4 nPa, *Bz*=0 nT, *Dst*=0 nT, *PS*=20°. The other model parameters are accepted as *By*=5 nT, G1=6 and G2=10.

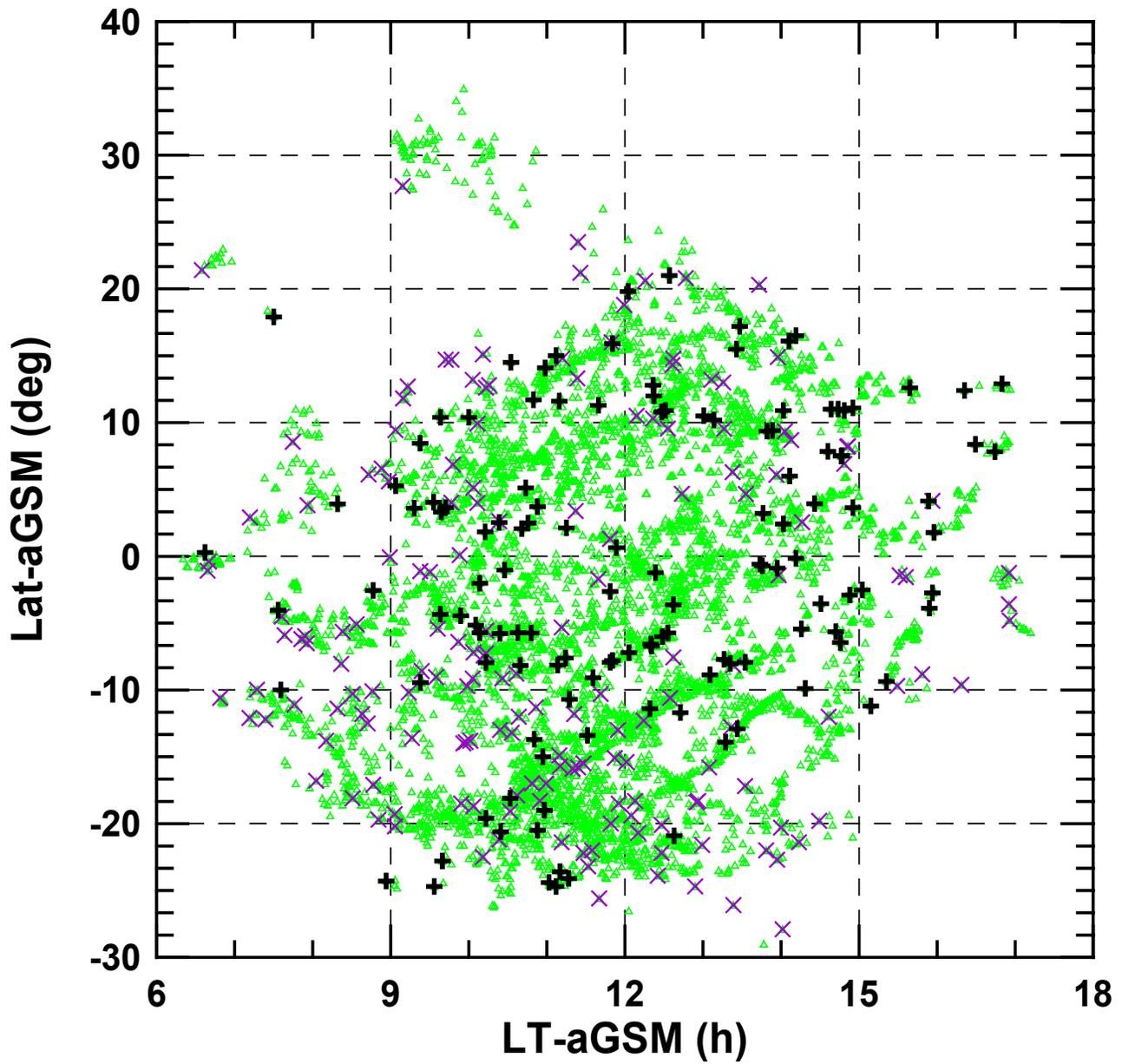

**Fig. 5.** Scatter plot of the GMCs in aGSM coordinates. Magnetopause crossings by GOES and LANL, and magnetosheath measurements are indicated by black crosses, gray crosses and triangles respectively.

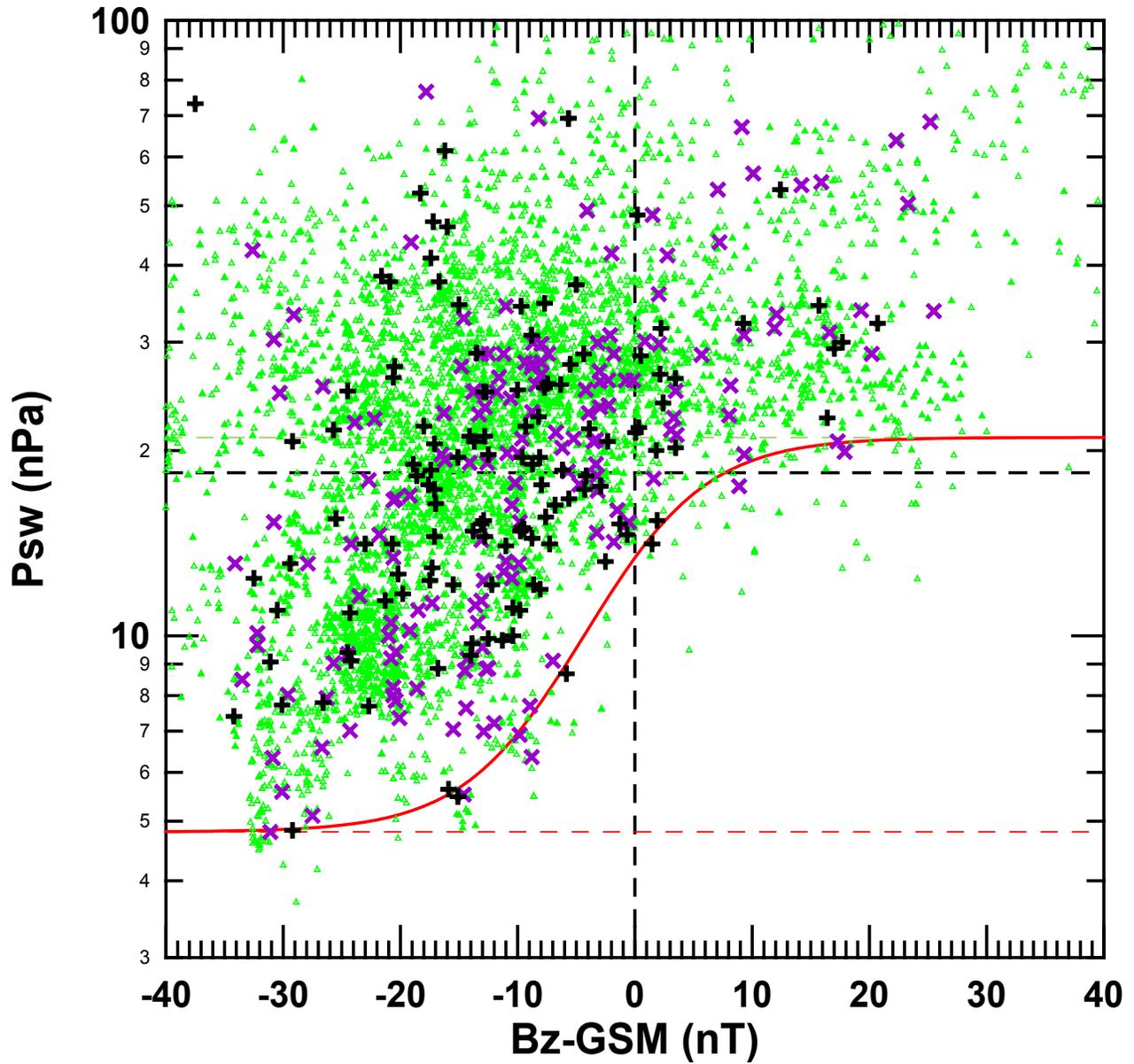

**Fig. 6.** Scatter plot Scatter plot of the GMCs in space of the solar wind parameters SW pressure versus aGSM *Bz*. Magnetopause crossings by GOES and LANL, and magnetosheath measurements are indicated by black crosses, gray crosses and triangles respectively. The thick curve is expressed by Eq. (4.3).

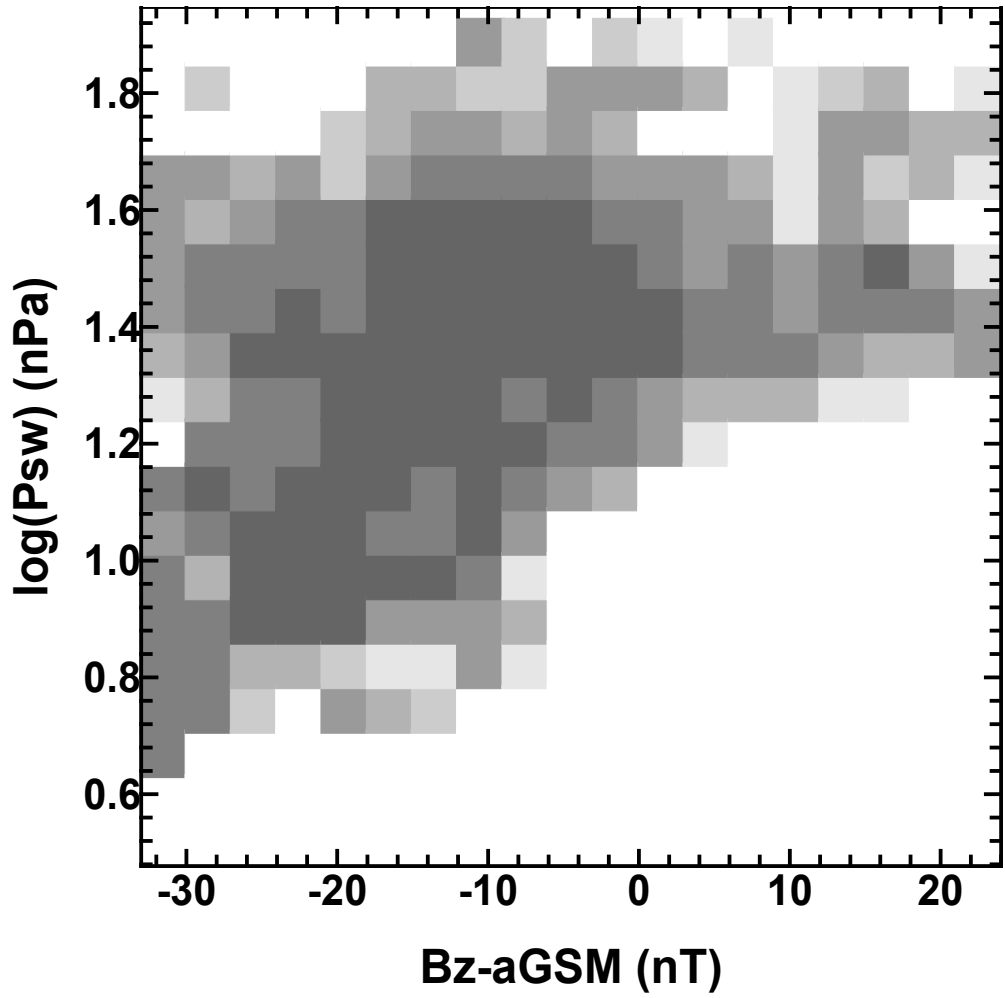

**Fig. 7.** Two-dimensional distribution of occurrence number of the GMC intervals binned in the coordinates *Psw* (in logarithmic scale) versus IMF *Bz* in aGSM.

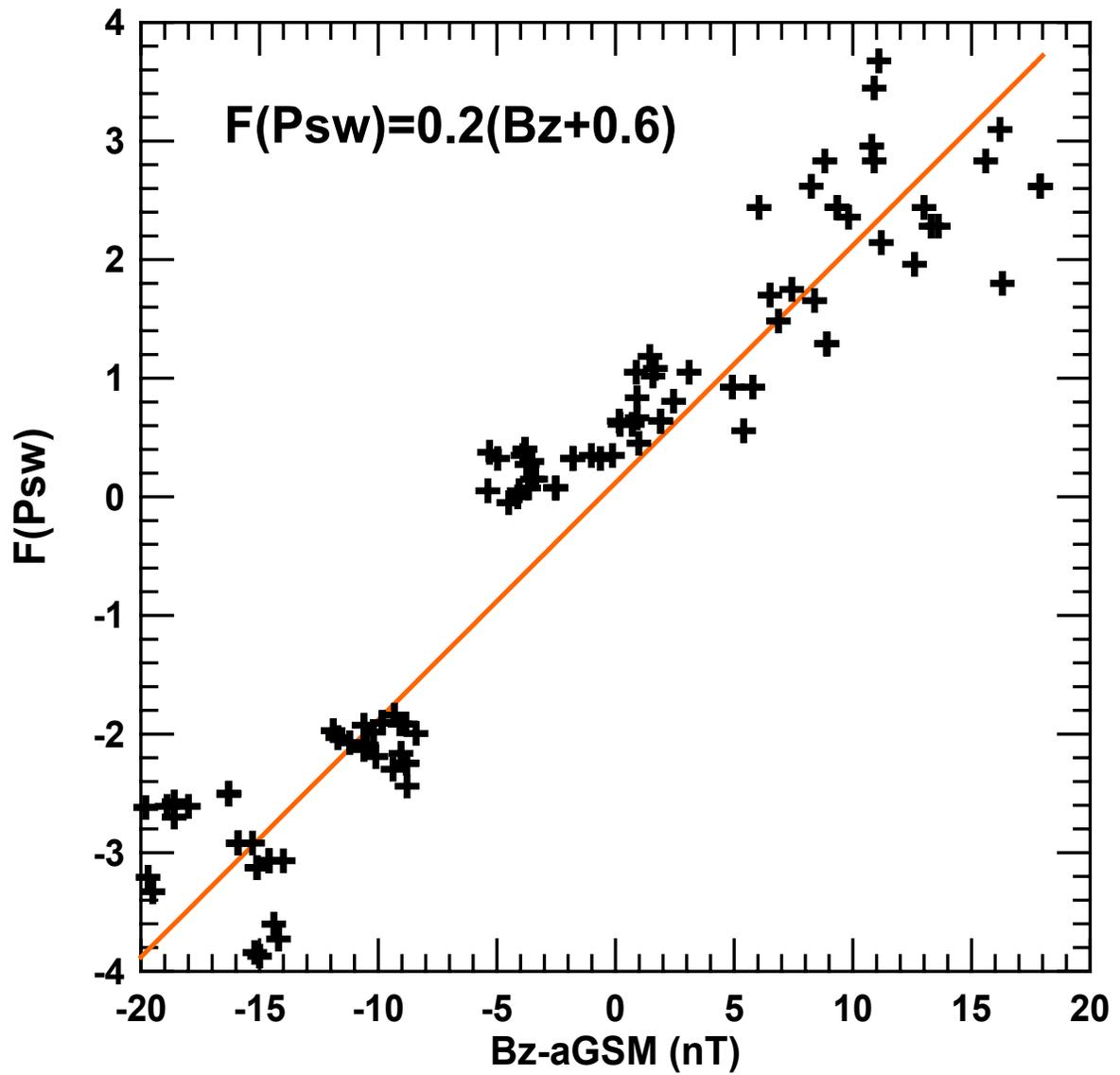

**Fig. 8.** Approximation of the meaningful GMCs and magnetosheath measurements from the down edge bins (Figure 6) by hyperbolic tangent in the space of parameters SW pressure and IMF *Bz*.

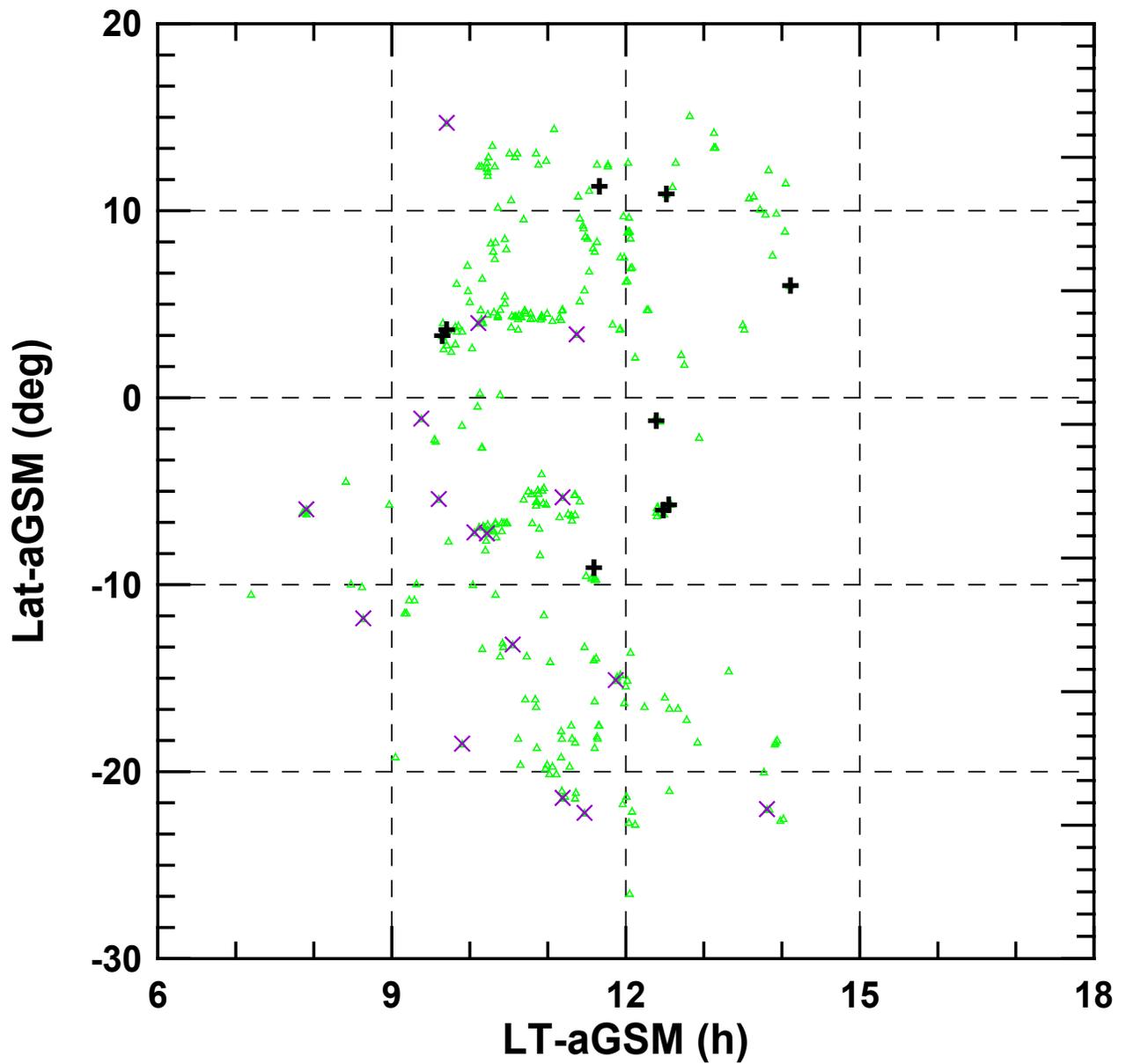

**Fig. 9.** Scatter plot of the magnetosheath measurements (triangles) and entrances by GOES (crosses) and LANL (asterisks) from 20% vicinity of the envelope boundary in aGSM coordinates latitude versus local time.

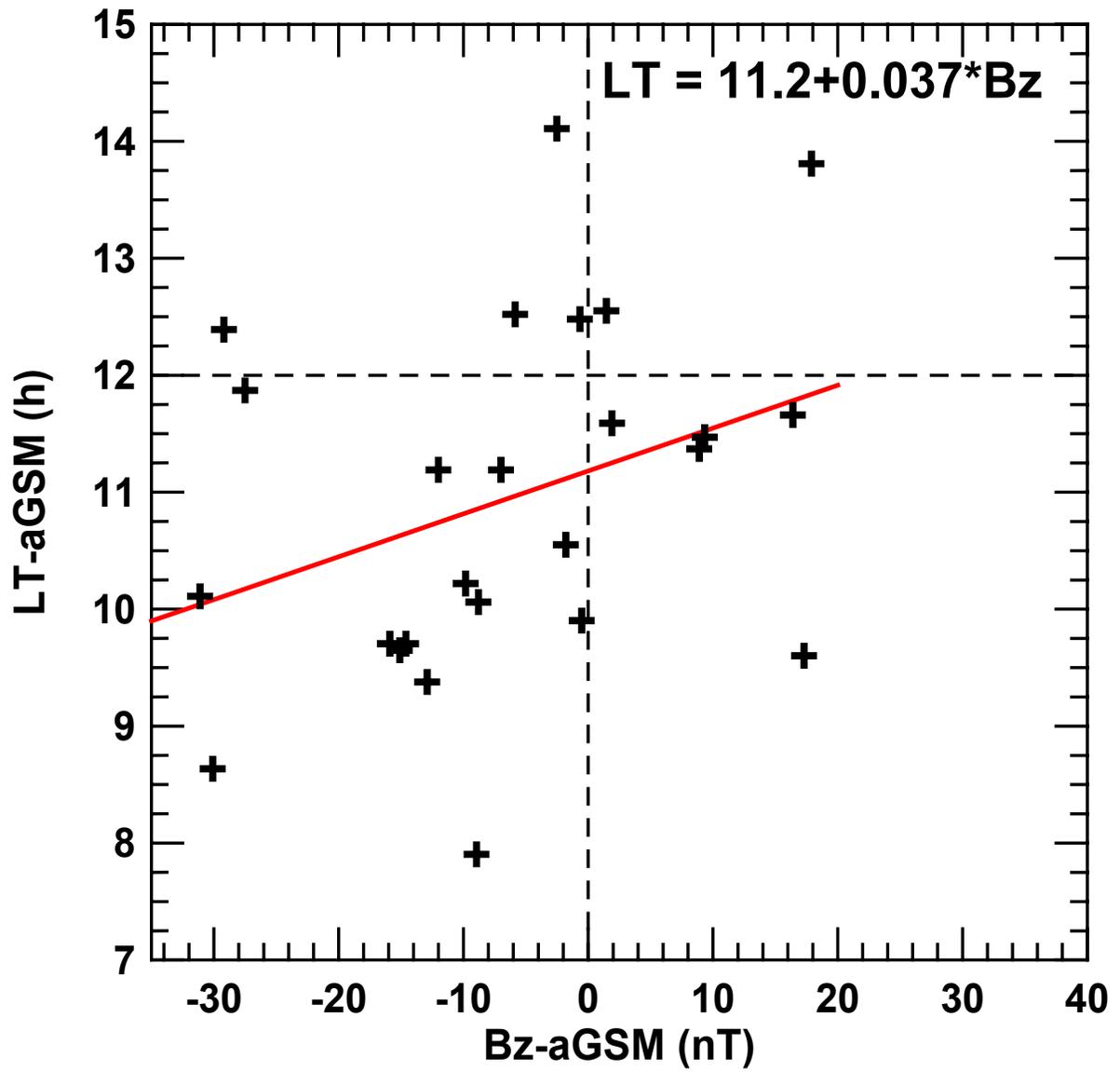

**Fig. 10.** Scatter plot of the magnetosheath entrances (crosses) from 20% vicinity of the envelope boundary in aGSM coordinates local time versus IMF *Bz*.

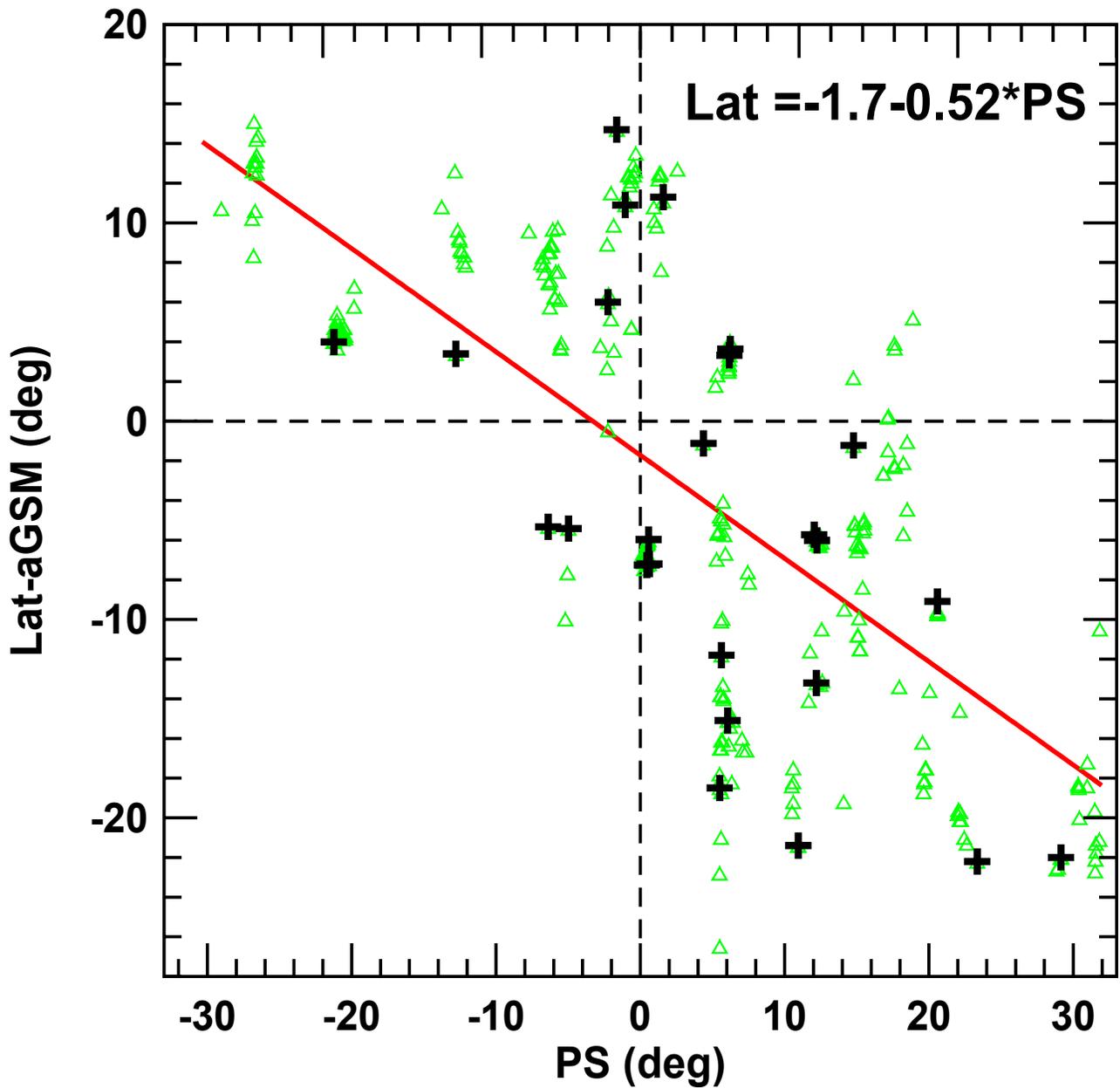

**Fig. 11.** Dependence of latitude of the GMCs (crosses) and magnetosheath measurements from 20% vicinity of the enveloping boundary on dipole tilt angle *PS*.

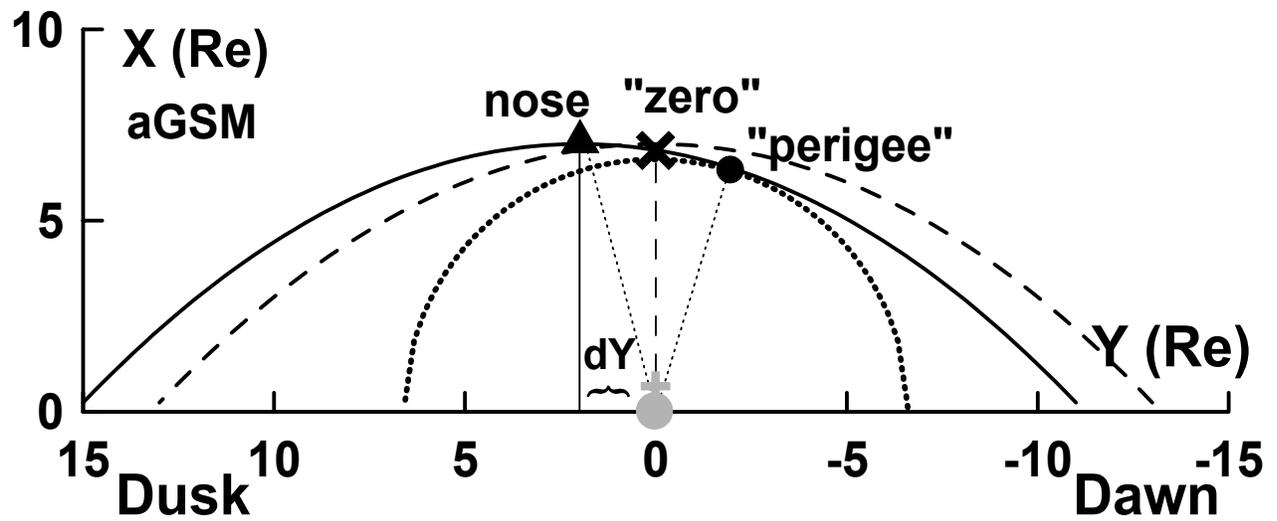

**Fig. 12.** KS98 model calculation for the magnetopause shifted on *dY*=2 Re toward the dusk (thick solid line) and "non-shifted" magnetopause (dashed line). The nose, "zero" and "perigee" points are indicated by circle, asterisk and triangle respectively.

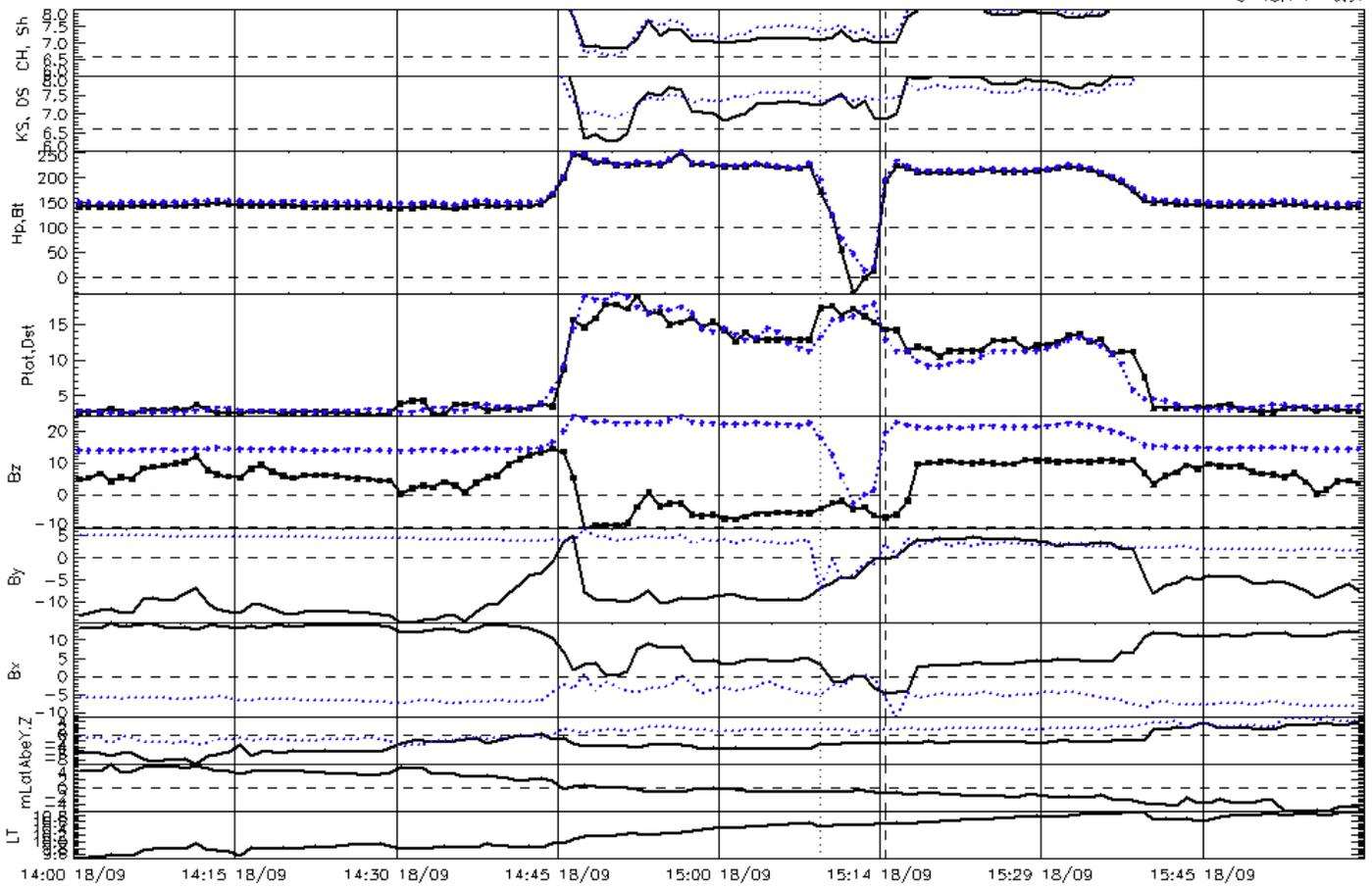

**Plot. 1.** GMC identification using GOES-8 magnetic measurements and solar wind data from the ACE upstream monitor on April 18, 2000. See details in the text.

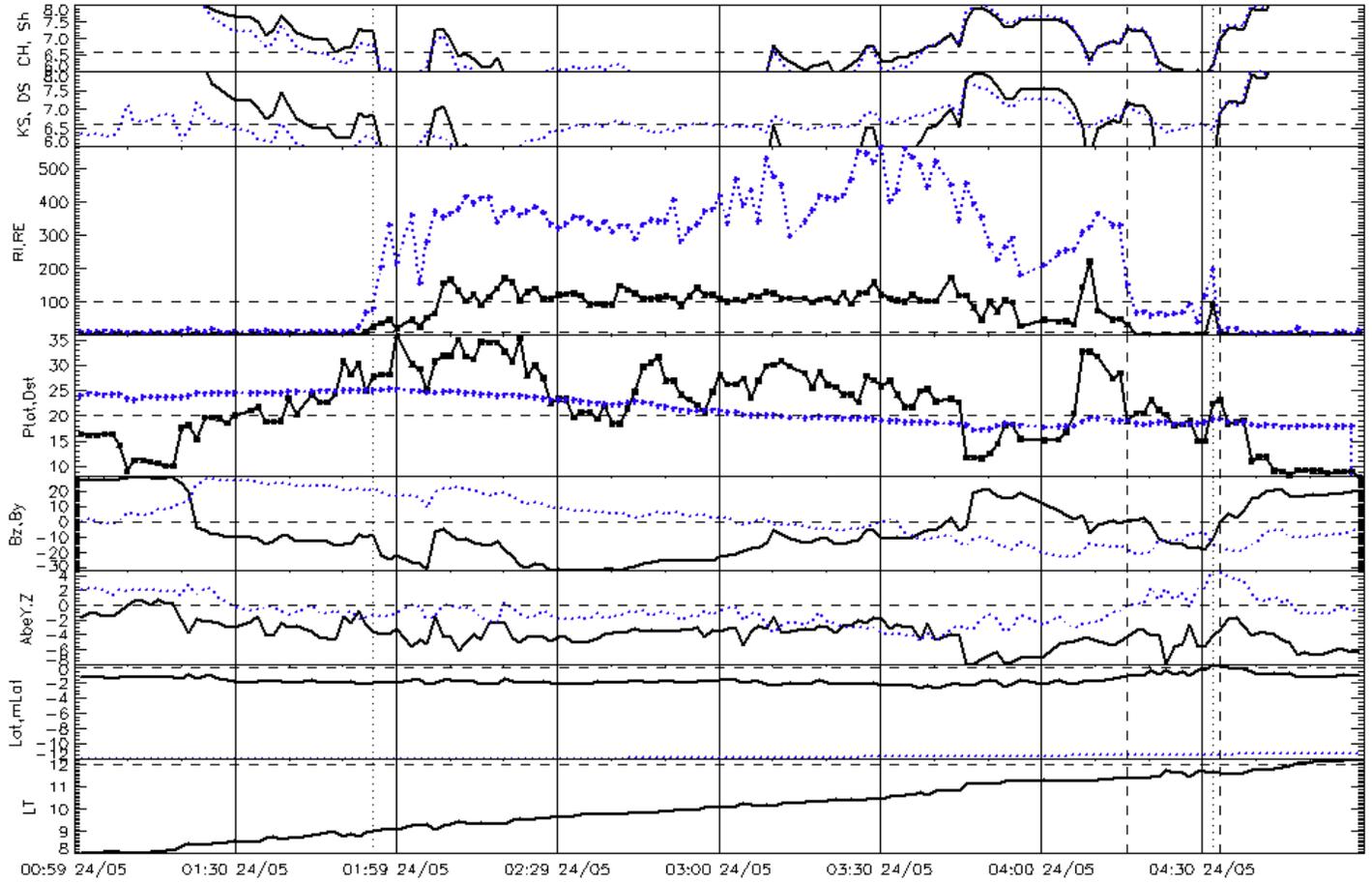

**Plot. 2.** GMC identification using LANL-1994 plasma measurements and solar wind data from the Wind upstream monitor on May 24, 2000. See details in the text.

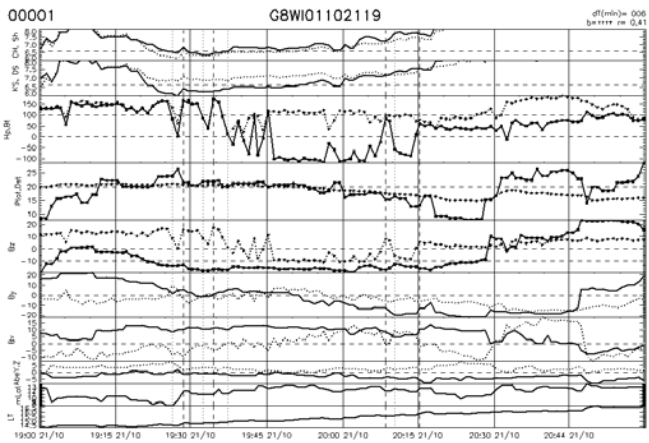
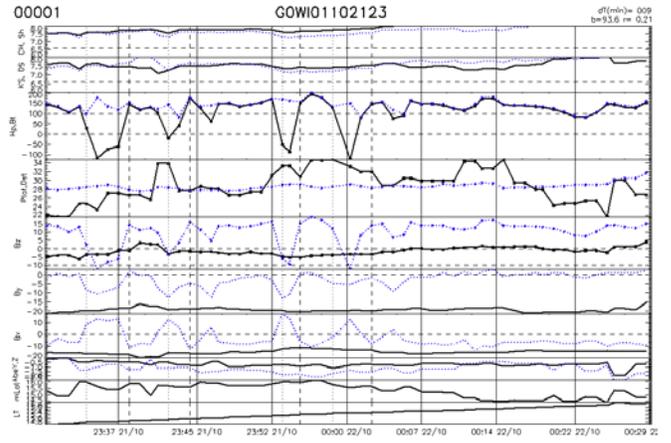
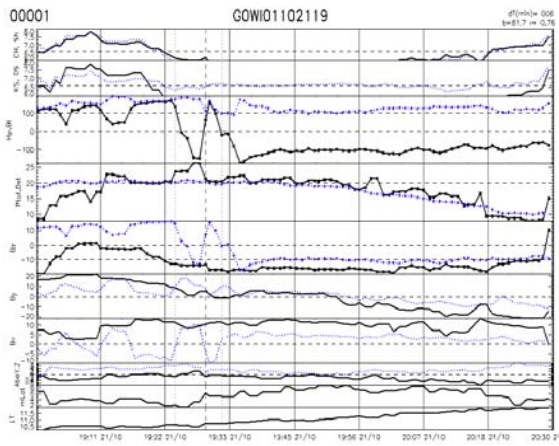
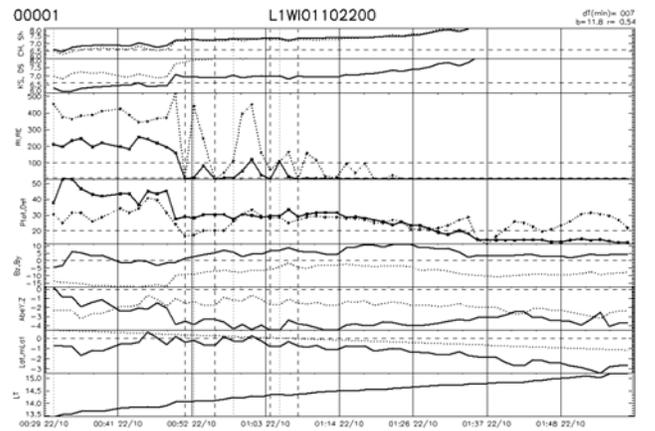

**Plot. 3.** Observations of high-amplitude surface waves on the magnetopause by GOES-8, 10 and LANL-1991 on October 21-22, 2001. See details in the text.

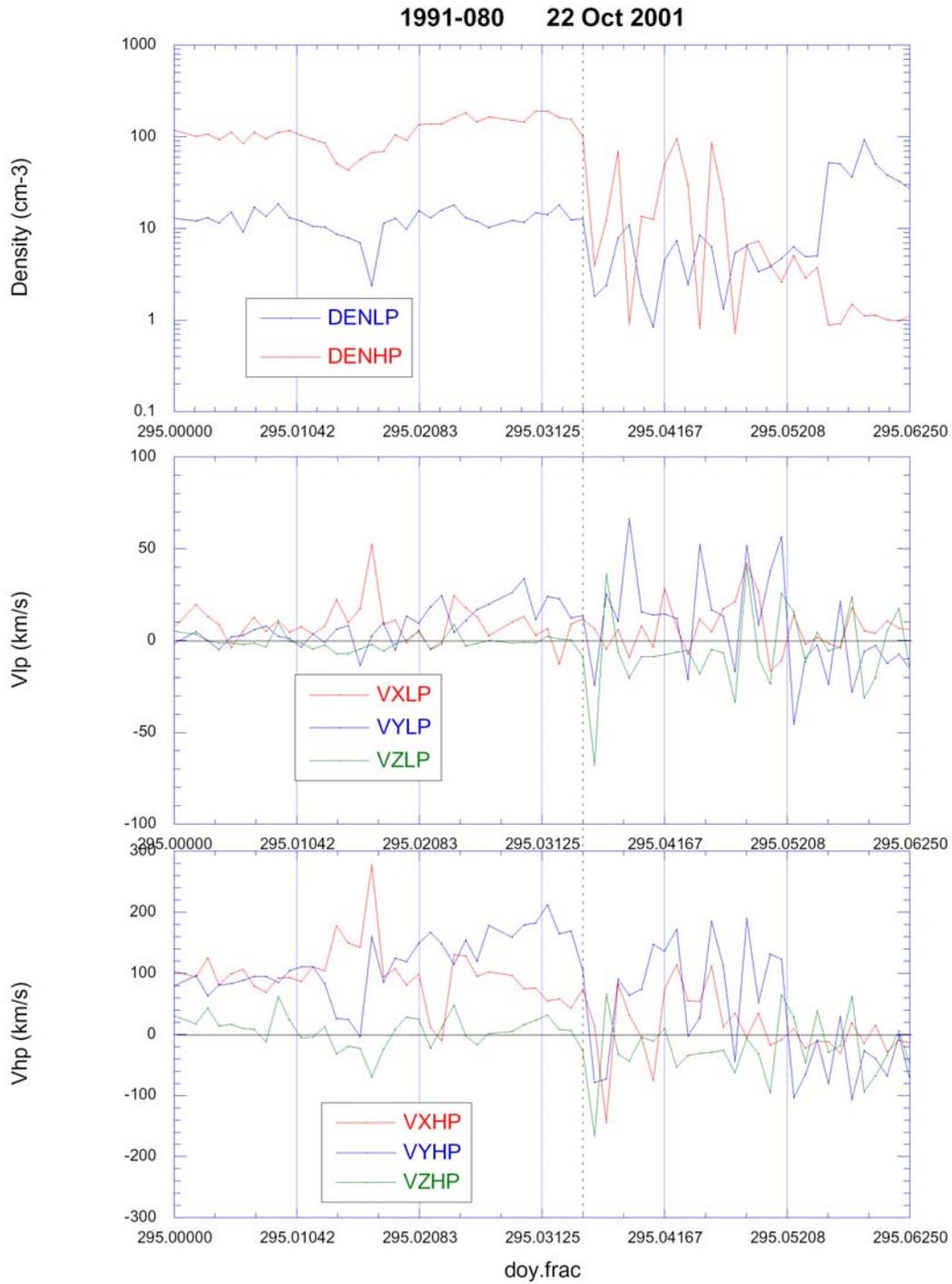

**Plot. 4.** LANL-1991 observations of high and low energy plasma (density and components of velocity) during high-amplitude surface waves on October 21-22, 2001. See details in the text.